\documentclass{aa}

\usepackage{graphicx}
\usepackage{txfonts}
\graphicspath{{figures/}{../figures/}}
\usepackage{caption, subcaption}
\usepackage{amsmath}
\usepackage{xcolor}
\usepackage{placeins}

\usepackage{hyperref}
\makeatletter
\renewcommand*\aa@pageof{, page \thepage{} of \pageref*{LastPage}}
\makeatother
\hypersetup{colorlinks=true,allcolors=[rgb]{0,0,0.8}}

\usepackage{soul}
\newcommand{\corr}[1]{#1}
\newcommand{\corrtwo}[1]{#1}

\begin{document}

\title{The e-MANTIS emulator: fast and accurate predictions of the halo mass function in $f(R)$CDM and $w$CDM cosmologies}
\titlerunning{HMF emulator in $f(R)$CDM and $w$CDM cosmologies}

\author{I. Sáez-Casares
	\inst{1}\thanks{inigo.saez-casares@obspm.fr}
    \and Y. Rasera \inst{1,4}
    \and T. R. G. Richardson \inst{2}
    \and P.-S. Corasaniti \inst{1,3}
}

\institute{Laboratoire Univers et Théories, Université Paris Cité, Observatoire de Paris, Université PSL, CNRS, F-92190 Meudon, France
\and 
Donostia International Physics Center (DIPC), Paseo Manuel de Lardizabal, 4, 20018, Donostia-San Sebasti\'an,
Gipuzkoa, Spain
\and
Sorbonne Universit\'e, CNRS, UMR 7095, Institut d'Astrophysique de Paris, 98 bis bd Arago, 75014 Paris, France
\and
Institut universitaire de France (IUF)\\
}



\abstract
{}
{
  In this work, we present a novel emulator of the halo mass function, which we implement in the framework of the \textsc{e-mantis} emulator of $f(R)$ gravity models.
  We also extend \textsc{e-mantis} to cover a larger cosmological parameter space and to include models of dark energy with a constant equation of state $w$CDM.
}
{
  We use a Latin hypercube sampling of the $w$CDM and $f(R)$CDM cosmological parameter spaces, over a wide range, and realize a large suite of more than $10000$ $N$-body simulations of different volume, mass resolution and random phase of the initial conditions. 
  For each simulation in the suite, we generate halo catalogues using the friends-of-friends (FoF) halo finder, as well as the spherical overdensity (SO) algorithm for different overdensity thresholds (200, 500 and 1000 times the critical density).
  We decompose the corresponding halo mass functions (HMF) \corrtwo{on} a B-spline basis, while adopting a minimal set of assumption on their shape.
  We use this decomposition to train an emulator based on Gaussian processes.
}
{
 The resulting emulator is able to predict the halo mass function for redshifts $\leq 1.5$ and for halo masses $M_h\geq10^{13}\,h^{-1}M_\odot$. 
 The typical HMF errors for SO haloes with $\Delta=200\mathrm{c}$ at $z=0$ in $w$CDM (respectively $f(R)$CDM) are of order of $\epsilon_0\simeq1.5\%$ ($\epsilon_0\simeq4\%$) up to a transition mass $M_t\simeq2\cdot10^{14}\,h^{-1}M_\odot$ ($M_t\simeq6\cdot10^{13}\,h^{-1}M_\odot$). For larger masses, the errors are dominated by shot-noise and scale as $\epsilon_0\cdot\left(M_h/M_t\right)^\alpha$ with  $\alpha\simeq0.9$ ($\alpha\simeq0.4$) up to $M_h \sim 10^{15}\,h^{-1}M_\odot$.
 Independently of this general trend, the emulator is able to provide an estimation of its own error as a function of the cosmological parameters, halo mass, and redshift.
 We have performed an extensive comparison against analytical parametrizations and shown that \textsc{e-mantis} is able to better capture the cosmological dependence of the HMF, while being complementary to other existing emulators.
}
{
  The \textsc{e-mantis} emulator, which is publicly available, can be used to obtain fast and accurate predictions of the halo mass function in the $f(R)$CDM and $w$CDM non-standard cosmological models.
  As such, it represents a useful theoretical tool to constrain the nature of dark energy using data from galaxy cluster surveys.
}


\maketitle
%
\section{Introduction}
\label{sec:intro}

Surveys of the large scale structures such as \corrtwo{the} \textit{Euclid} \corrtwo{mission} \citep{2018LRR....21....2A}, \corrtwo{the Dark Energy Spectroscopic Instrument (DESI) experiment} \citep{2016arXiv161100036D} or the Legacy Survey of Space and Time (LSST) at the \textit{Vera Rubin} \corrtwo{Observatory} \citep{2019ApJ...873..111I} will map the distribution of galaxies and probe the clustering of matter in the universe over an unprecedented range of scales and redshifts.
There is a wide-spread consensus that these observations can provide insights on the nature of dark energy and test the validity of General Relativity on cosmic scales \citep[see e.g.][for a review]{2016ARNPS..66...95J}.
However, this requires theoretical model predictions capable of accounting for the effects of the non-linear gravitational collapse of matter at late time and at small scales.
Unlike linear perturbation theory, accurate predictions of such effects require the use of large volume high-resolution $N$-body simulations.
Nevertheless, a single simulation can only provide predictions for a given cosmological model and more specifically the set of cosmological parameters assumed in the simulation.
Given their large computational cost, this poses a challenge to future cosmological parameter inference analyses of large scale structure data. 

Different approaches have been developed to address this issue.
Semi-analytical methods, relying on a combination of theoretical modelling and calibration to simulations, have been formulated to compute the non-linear matter power spectrum  \citep[see e.g.][]{2003MNRAS.341.1311S,2012ApJ...761..152T,2016JCAP...03..057V,2017JCAP...04..026C,2023MNRAS.519.4780B} or the halo mass function \citep[see e.g.][]{ST1999,Tinker_2008,Castro2021}.
These have the advantage of providing cosmological model predictions at negligible computational costs.
Alternatively, approximate numerical schemes have been introduced to perform simplified fast numerical simulations \citep[e.g.][]{2002MNRAS.331..587M,2013JCAP...06..036T}.
However, these approaches all come with their own limitations and uncertainties that should be propagated in cosmological data analyses.
In the last decade, the use of machine-learning algorithms have led to a third way based on the concept of emulators, initially proposed in the seminal work by \citet{habib_emulator}.
Their use in large scale structure data analyses has become commonplace \citep[e.g.][]{euclid_emulator, euclid_emulator_2, Angulo2021, Moran2023, Storey-Fisher_2024}.
In fact, cosmological emulators have proven to be very promising tools, as in theory they can be arbitrarily precise and incorporate as many effects as their training dataset provides.
While this increase in precision may come at a higher computational cost of generating the training dataset, the resulting emulator is able to provide extremely fast and accurate predictions, which gives them excellent performance when placed within a likelihood designed for a Markov Chain  Monte Carlo (MCMC) analysis.

In this work, we present an emulator of the halo mass function (HMF), which we implement within the wider framework of the \textsc{e-mantis}\footnote{Emulator for Multiple observable ANalysis in extended cosmological TheorIeS}\footnote{\url{https://zenodo.org/doi/10.5281/zenodo.7738362}}\footnote{\url{https://gitlab.obspm.fr/e-mantis/e-mantis}} emulator of $f(R)$CDM models \citep{emantis_ps_boost}, that we extend to a larger cosmological parameter space and include models of dark energy with constant equation of state $w$CDM. 
In recent years, the increasing availability of simulations of larger volumes and higher resolution has led to parametrizations of the HMF providing evermore precise fits to simulations, capable of providing better descriptions of the mass, redshift, and cosmological parameter dependence \citep[e.g.][]{Jenkins2001, Crocce_2010, Courtin_2011, Angulo_2012, Watson_2013,Despali_2016,Bocquet_2016,Diemer2020,Ondaro-Mallea_2022,Castro2021}.
These parametrizations have since been used as a core part of cosmological studies, such as cluster number count analyses \citep{Allen2011, Kravtsov2012, Rozo2010, Mantz2015, de_Haan2016, Planck2016_clusters, Schellenberger2017, Pacaud2018, Bocquet2019, Abbott2020, DESY1_clusters, Lesci2022}, which are rapidly gaining traction due to the increasing size of cluster samples the community has, and will have, access too \citep[e.g.][]{Hofmann2017, Euclid_cluster_detection_forecast, Raghunathan2022}.
Nonetheless, in the current era of precision cosmology, several of the previous fitting functions have become insufficiently accurate given survey and mission requirements \citep[see e.g.][]{Bocquet_2016, Castro2021} or are simply unable to explore the wider parameter space of cosmological models considered.
This motivates the development of dedicated emulators capable of accurately predicting the HMF for different cosmological scenarios and cosmological parameter values.
A number of emulators of the HMF in $w$CDM models have already been developed in the literature \citep[see e.g.][]{dark_quest_hmf_emu, aemulus_hmf_emu, mira_titan_hmf_emu}.
However, some of them \citep[such as][]{aemulus_hmf_emu}, are limited to a narrow emulation range around current observational constraints.
Additionally, existing $w$CDM HMF emulators usually only support a single dark matter halo definition, which is not necessarily the same as the ones used in usual observational analyses.
In the case of $f(R)$CDM cosmology, the only existing emulator for the HMF is the one presented in \citet{forge_hmf_emu}.
This emulator does not cover the mass range of galaxy clusters, since it is aimed at modelling galaxy clustering.
Here, we present a novel emulator, that supports multiple dark matter halo definitions commonly used in both theoretical and observational studies.
It covers a wider parameter space than existing $w$CDM emulators, and it is the first emulator for the $f(R)$CDM cosmologies in the galaxy cluster mass range.

This paper is structured as follows.
Section~\ref{subsec:emantis_sims} presents the extended \textsc{e-mantis} $N$-body simulation suite.
Section~\ref{subsec:emulating_hmf} explains the methodology employed to emulate the HMF.
In Sect.~\ref{sec:results} we validate the predictions of the emulator and compare them to external simulations and other publicly available predictions.

\section{The extended e-MANTIS simulation suite}
\label{subsec:emantis_sims}

\subsection{Alternative dark energy models}\label{subsec:de_models}

In the standard cosmological scenario, the cosmological constant $\Lambda$ is equivalent to a homogeneous static fluid characterized by an equation of state $w=-1$ corresponding to a constant energy density.
By contrast, a dynamical form of dark energy can be described as a fluid with a constant equation of state $w\ne -1$ provided that $w<-1/3$, such as to ensure it sources the cosmic acceleration.
In such a case, the dark energy density evolves as a power-law function of redshift, i.e. $\rho_{\rm DE}\propto (1+z)^{3(1+w)}$.
Models with $-1<w<-1/3$ can approximate the dynamics of quintessence scalar field scenarios \citep{1998PhRvL..80.1582C}, while models with $w<-1$ corresponds to the so-called phantom dark energy scenario \citep{2003PhRvL..91g1301C}, that can also mimic unaccounted scalar interactions between dark energy and dark matter \citep{2006PhRvD..73h3509D}.
As such, the $w$CDM model is a simpler realization of dynamical dark energy scenarios.

Alternatively, dark energy may result from a modification of Einstein's theory of General Relativity (GR) on cosmic scales \citep[see e.g.][]{Clifton_2012}.
One popular example of modified gravity (MG) is the $f(R)$ gravity model \citep[see e.g.][]{Sotiriou_2010}.
This takes the form of a modification of the Einstein-Hilbert action by the addition of a function $f$ of the Ricci scalar $R$:
\begin{equation}
S_{g} = \frac{1}{2\kappa^2} \int \mathrm{d}^4x\sqrt{-g}\left[R+f(R)\right], 
\end{equation}
with $\kappa^2=8\pi G/c^4$ where $G$ is Newton's constant, $c$ the speed of light in vacuum and $g$ the determinant of the space-time metric $g_{\mu\nu}$.
Here, we focus on a specific $f(R)$ gravity model proposed by \citet{2007PhRvD..76f4004H}, which assumes
\begin{equation}
f(R)=-m^2\frac{c_1 \left(R/m^2\right)^n}{c_2\left(R/m^2\right)^n+1},
\end{equation}
where $c_1$, $c_2$ and $n$ are dimensionless parameters and $m$ is a curvature
scale.
The latter is set to 
\begin{equation}
    m^2=\frac{\Omega_{\rm m}\,H_0^2}{c^2},
\end{equation}
where $\Omega_{\rm m}$ is the cosmic matter density today and $H_0$ the Hubble rate.
The coefficient $c_1$ and $c_2$ can be set to
\begin{equation}
    \frac{c_1}{c_2}=6\frac{\Omega_{\Lambda}}{\Omega_{\rm m}},
\end{equation}
such that in the high curvature limit ($R\gg m^2$) the model matches the accelerating phase of the $\Lambda$CDM model.
Hence, in such a limit, the function $f$ reads as 
\begin{equation}
    f(R)\approx -2\Lambda +\frac{f_{R_0}}{n} \frac{R_0^{n+1}}{R^n},
    \label{eq:HS_approx}
\end{equation}
where $f_{R_0}$ is the derivative of the scalar function with respect to the Ricci scalar evaluated at $z=0$.
At the level of the background expansion, $f_{R_0}$ is a parameter that sets the deviation with respect to $w=-1$ of the cosmological constant case \citep{2007PhRvD..76f4004H}.
Here, we consider $f(R)$ gravity models characterized by deviations from GR no greater than $f_{R_0}=-10^{-4}$.
From a phenomenological point of view, these models have a background expansion indistinguishable from that of the $\Lambda$CDM model, with negligible differences concerning the evolution of the matter density perturbations at high redshift.
On the other hand, differences arise in the late-time non-linear regime of cosmic structure formation, thus requiring the use of numerical simulations.

Here, we consider $f(R)$CDM models with $n=1$.
We remark that an emulator of the matter power spectrum for models with $n\neq 1$ has been developed using an approximate $N$-body solver \citep{ramachandra_emulator}.
This is because $N$-body simulations of $f(R)$ models with $n\neq1$ are significantly more time-consuming than the case $n=1$ ~\citep[see][]{Bose2017}.
However, this approach is not sufficiently accurate at the level we wish to capture on non-linear scales.
Henceforth, as a starting point to build an accurate and efficient emulator of $f(R)$ gravity models, we focus on the case $n=1$ and we leave the extension to $n\neq 1$ to future work.
We note that a simple particle-mesh solver has been developed by \citet{Ruan_2022}, to simulate the non-linear clustering of matter in the cases $n=0$ and $n=2$, which also benefit from the same speed-up as $n=1$.

\subsection{Cosmological parameter space}
\label{subsec:param_space_range}

\begin{table*}
  \caption{
    Ranges of the cosmological parameters supported by the emulator, and values of the cosmological parameters for some reference models.
  }
	\label{table:cosmo_params}
	\centering
	\begin{tabular}{c c c c c c c c}
		\hline\hline
		Model & $w$ &  $\log_{10} \left|f_{R_0}\right|$ & $\Omega_{\rm m}$ & $\sigma_8^{\rm GR}$ & $h$ & $n_{\rm s}$ & $\Omega_{\rm b}$ \\
		\hline
        $w$CDM & $\left[-1.5, -0.5\right]$ & - & $\left[0.155, 0.465\right]$ & $\left[0.6083, 1.014\right]$ & $\left[0.55, 0.85\right]$ & $\left[0.72, 1.2\right]$ & $\left[0.037, 0.062\right]$ \\
        $f(R)$CDM & - & $\left[-7, -4\right]$ & $\left[0.155, 0.465\right]$ & $\left[0.6083, 1.014\right]$ & $\left[0.55, 0.85\right]$ & $\left[0.72, 1.2\right]$ & $\left[0.037, 0.062\right]$ \\
        \hline
        P$18$ & $-1$ & - & $0.3071$ & $0.8224$ & $0.6803$ & $0.96641$ & $0.048446$ \\
        w$12$ & $-1.2$ & - & $0.3071$ & $0.8224$ & $0.6803$ & $0.96641$ & $0.048446$ \\
        F$5$ & - & $-5$ & $0.3071$ & $0.8224$ & $0.6803$ & $0.96641$ & $0.048446$ \\
        F$6$ & - & $-6$ & $0.3071$ & $0.8224$ & $0.6803$ & $0.96641$ & $0.048446$ \\
	\end{tabular}
    \tablefoot{
      Model is the cosmology, $w$ the equation of state of dark energy, $f_{R_0}$ the derivative of the scalar function with respect to the Ricci scalar at $z=0$, $\Omega_{\rm m}$ the matter density parameter,  $\sigma_8^{\rm GR}$ the amplitude of linear matter density fluctuations on the $8\,h^{-1}\mathrm{Mpc}$ scale assuming GR, $h$ the normalized \corr{Hubble parameter}, $n_{\rm s}$ the slope of the primordial power spectrum, $\Omega_{\rm b}$ the baryon density parameter.
    }
\end{table*}

In the case of the $w$CDM scenario, dark energy is specified by the value of a constant equation of state parameter, $w$.
In the case of the $f(R)$CDM scenario, the amplitude of the gravity modifications relative to GR is specified by the value of $f_{R_0}$.
The rest of the parameter space for both scenarios is specified by the values of the standard cosmological parameters: the total matter density $\Omega_{\mathrm{m}}$, the scalar spectral index $n_s$, the reduced Hubble rate $h$\footnote{Defined as $H_{0} \equiv 100 \, h\,\mathrm{km}\,\mathrm{s}^{-1}\mathrm{Mpc}^{-1}$, where $H_{0}$ is the Hubble rate today.}, the baryon density parameter $\Omega_{\mathrm{b}}$ and the amplitude of linear matter density fluctuations on the $8\,h^{-1}\mathrm{Mpc}$ scale $\sigma_8^{\rm GR}$ in GR.
By $\sigma_{8}^{\rm GR}$, we refer to the $\sigma_{8}$ obtained assuming a linear GR evolution.
In the case of $w$CDM, this corresponds identically to the usual $\sigma_{8}$.
In the case of $f(R)$CDM, we follow the usual convention where $\sigma_{8}^{\rm GR}$ corresponds to the value of $\sigma_{8}$ in $\Lambda$CDM with identical cosmological parameters other than $f_{R_{0}}$.
We assume a flat geometry, we ignore massive neutrinos and set the radiation density parameter to the value given by the \corrtwo{cosmic microwave background (CMB)} temperature, $\Omega_{\mathrm{r}}\sim10^{-4}$.

It is worth stressing that the predictions from emulators are limited to the range of parameters initially selected to generate the training set.
Extending the emulator to a wider range or to support additional parameters is not straightforward, as it usually requires restarting the whole procedure.
Because of this, it is important for an emulator to have a sufficiently large parameter space (larger than current cosmological constraints) to be effectively used for observational data analyses.
A large range is also needed in order to capture the subtle deviations from universality of the dark matter halo mass function \citep{Tinker_2008,Courtin_2011, Ondaro-Mallea_2022}.
On the other hand, considering a too large parameter space may lead to a lower density of training points and consequently degrade the final accuracy of the emulator.
Guided by the choice made in previous emulators of $w$CDM and $f(R)$CDM models, we consider the following range of values for $w$ and $f_{R_0}$ respectively: $w\in[-1.5,-0.5]$ and $\log_\mathrm{10}{|f_{R_{0}}|}\in [-7,-4]$; while we set the range of the other cosmological parameters to the following values: $\Omega_\mathrm{m}\in[0.155,0.465]$, $\sigma_8^{\rm GR}\in[0.608,1.014]$, $h\in[0.55,0.85]$, $n_s\in[0.72,1.20]$ and $\Omega_{\rm b}\in [0.037,0.064]$.
These values are summarized in Table~\ref{table:cosmo_params}.

The ranges we consider are wider than those assumed in existing emulators of the HMF for $w$CDM models \citep[see e.g.][]{aemulus_hmf_emu, dark_quest_hmf_emu}.
This is also the case of the \textsc{mira-titan} emulator~\citep{mira_titan_hmf_emu}, for which we consider a larger range of values of $w$.
On the other hand, the \textsc{mira-titan} emulator supports the case of massive neutrinos and a time-varying equation of state, which we do not include here.

As far as the emulators of $f(R)$CDM models are concerned, the $N$-body simulation suite used here is the first to fully explore the $6$-dimensional cosmological parameter space of such models.
As an example, the \textsc{forge} $N$-body simulation suite~\citep{Arnold2021} does not include variations of the scalar spectral index $n_{\rm s}$ and the \corr{baryon density parameter} $\Omega_{\rm b}$, and has a more restricted range in terms of the $f_{R_0}$ parameter.
The approximate simulations presented in \citet{ramachandra_emulator} do not take into account the reduced Hubble constant $h$, although they consider variations of the $n$ parameter.
In this work, we extend the base of parameters sampled in the first version of the \textsc{e-mantis} emulator presented in \citet{emantis_ps_boost}, to include the dependence on $n_{\mathrm{s}}$, $h$ and $\Omega_{\mathrm{b}}$, while sampling a wider interval of $\Omega_{\mathrm{m}}$ values.

\subsection{Sampling the parameter space}
\label{subsec:param_space_sampling}

A common approach to efficiently sample the multidimensional cosmological parameter space with a reduced number of $N$-body simulation consists in using a Latin Hypercube Sampling (LHS) method~\citep{McKay_1979}.
Such a strategy has been shown to provide accurate emulators when combined with Gaussian processes (GP)~\citep{habib_emulator}.

In order to sample a parameter space with $N$ points, the LHS method first divides the $D$-dimensional parameter space with a $N^{D}$ regular grid.
Then, $N$ random cells are selected, with the constraint that any two cells cannot have any common coordinates. The space filling properties of the LHS can be further improved by applying different optimizations, such as maximizing the minimum distance between points (maximin-distance condition).
One limitation of the simple LHS approach, is that once a sample of $N$ points has been selected, there is no direct way of adding new points.
In practice, it is usually difficult to know in advance the exact number of simulations necessary to achieve a satisfactory level of emulation accuracy.
One possibility is to use approximate analytical models as surrogate training data to determine the number of cosmological models which need to be sampled.
However, this would only apply to a given particular observable.

In the work of~\citet{plhs}, the authors have introduced a method to generate a Progressive Latin Hypercube Sampling (PLHS).
A PLHS generates multiple sub-sets (slices) of points, such that each slice is a LHS and the progressive union of all slices remains a LHS at all stages.
It is possible to generate as many new slices as desired.
However, building a PLHS requires doubling the number of points with each new slice.
This procedure becomes impractical in the context of expensive numerical simulations and does not allow a fine control of the final number of training models. These issues can be somewhat alleviated through the use of a quasi-PLHS, which we use to construct our training data set. As such, we briefly introduce the main aspects of the quasi-PLHS method below, but refer the interested reader to \citet{plhs} for more details.

The quasi-PLHS method is based on the Sliced Latin Hypercube Sampling (SLHS) method.
A SLHS is an ensemble of a finite number of LHS slices, the union of which also forms a LHS.
However, the progressive union of slices does not constitute a LHS at each step, as would be the case with a perfect PLHS.
An example of a method to generate an SLHS is presented in~\citet{LHS_sliced}.
The main idea behind the quasi-PLHS is to find an optimum ordering of the slices, so that their progressive union is as close as possible to a LHS.
Using this method, it is possible to progressively increase the number of points in the sample with an arbitrary constant step up to a maximum number of points fixed in advance.
By fixing the maximum number of points to a sufficiently large value, it is possible to add as many models as needed to the training samples progressively.
The price to pay for a more flexible sampling method, is that for a given number of slices, the distribution of points is less optimal than a one-shot LHS of the same size.
The only cases where we obtain a real LHS, is either by using a single slice or all of them.

We generate a maximin-distance SLHS following the method introduced by~\citet{LHS_sliced}.
The optimized distance is a combination of the distance between points in each slice and in the union of all slices.
We use the R package \textsc{slhd}\footnote{https://cran.r-project.org/web/packages/SLHD/index.html} to generate $16$ slices with $16$ points each, resulting in $256$ models.
The total number of $256$ models is a conservative upper limit, since most cosmological $N$-body based emulators do not use that many training simulations.
We then use the quasi-PLHS method of~\citet{plhs} to find an optimum ordering of the slices.
At the current stage, we have run $N$-body simulations for the first $5$ slices (i.e. $80$ models).
The other $11$ slices remain available for future extension of the simulation suite.
Figure~\ref{fig:LHS} shows the final distribution of cosmological models in both $f(R)$CDM and $w$CDM.
In both models, we use exactly the same values of $\Omega_{\mathrm{m}}, \sigma_{8}^{\rm GR}, n_{\mathrm{s}}, h$ and $\Omega_{\mathrm{b}}$.
For $w$ and $f_{R_0}$ we use the same distribution of points, which is simply scaled to their respective range of values.

\subsection{Simulation pipeline}
\label{subsec:sim_pipeline}

The simulation suite presented in this work has been run using the automated simulation pipeline described in \citet{emantis_ps_boost}, which is an extension to $f(R)$ gravity models of the original $w$CDM pipeline developed for the Parallel-Universe-Runs (PUR) project \citep{Blot2015,Blot2021}.
Compared to the original version of the simulation pipeline, several improvements have been implemented.
We describe here the updated pipeline used in this work.

The linear matter power spectra necessary to the generation of initial conditions were computed using \textsc{camb} \citep{Lewis2000}.
The initial conditions, particle displacements and velocities, were generated using a modified version of the code \textsc{mpgrafic}~\citep{Prunet2008}, which computes the displacement field with second-order Lagrangian perturbation theory. The starting redshift was taken as late as possible while avoiding particle crossing (i.e. $z_{i}\sim50$) in order to minimize discreteness errors \citep{Michaux_2021}. 
The dark matter particles have been evolved with the code \textsc{ecosmog} \citep{Li2012,Bose2017}, a modified gravity version of the Adaptive Mesh Refinement $N$-body code \textsc{ramses} \citep{Teyssier2002,Guillet2011}.
We have modified \textsc{ecosmog} in order to solve the Poisson equation for the sum of the two Bardeen potentials ($\phi$ and $\psi$), which is relevant for future weak-lensing studies.
Finally, the matter power spectra have been computed with the code \textsc{powergrid} \citep{Prunet2008}.

Haloes were detected using two distinct halo finder codes: \textsc{pfof}, a parallelized version of the Friends-of-Friends (FoF) percolation algorithm described in \citet{roy2014pfof};
\textsc{psod}, a parallelized version of the  Spherical Overdensity (SO) algorithm \citep{LaceyCole94_SO} based on the \textsc{pfof} parallel scheme.
In the case of the SO halo finder, the Cloud-In-Cell density of all particles is computed on a grid twice thinner than the coarse grid of the dynamical solver.
Only “dense” particles with density larger than 120 times the matter density are selected as possible candidates for halo centres.
We note that decreasing this threshold does not change the final result.
The selected particles are then considered in decreasing density order.
For each particle considered, we scan all particles which belong to the same coarse cell and those in the 124 neighbouring cells to find the particle with the lowest value of the gravitational potential, as computed by \textsc{ecosmog}.
In previous versions, the centre was identified as the position of the densest particle or the location of the minimum of the self-binding energy.
Once the centre is found, a sphere of increasing radius is grown outwards until the average density within the sphere, including all particles, drops below a given density threshold.
Finally, all the particles within the sphere are removed from the list of possible centres and the process is repeated.
We store the particles of each detected halo, thus allowing us to compute various halo properties in a post-processing step. 

\subsection{The \texorpdfstring{$N$}{N}-body simulation suite}
\label{subsec:nbody_suite}

\begin{table*}
  \caption{
    Summary of the main characteristics of the different simulation sets used in this work.
  }
	\label{table:sim_suite}
	\centering
	\begin{tabular}{c c c c c c c}
		\hline\hline
		Name & Model(s) &  $L_{\mathrm{box}} \, [h^{-1}\mathrm{Mpc}]$ & $N_{\mathrm{part}}$ & $m_{\mathrm{part}} \, [\left(\Omega_{\rm m}/0.3071\right)h^{-1}M_{\odot}]$ & $N_{\mathrm{real}}$ & $N_{\mathrm{cosmo}}$ \\
		\hline
		$\mathrm{L}328\_\mathrm{M}10\_\mathrm{wcdm}$ & $w$CDM & $328.125$ & $512^{3}$ & $2.25\cdot10^{10}$ & $64$ & $80$ \\
		$\mathrm{L}656\_\mathrm{M}11\_\mathrm{wcdm}$ & $w$CDM & $656.25$ & $512^{3}$ & $1.79\cdot10^{11}$ & $64$ & $80$ \\
		$\mathrm{L}328\_\mathrm{M}10\_\mathrm{frcdm}$ & $f(R)$CDM & $328.125$ & $512^{3}$ & $2.25\cdot10^{10}$ & $8$ & $80$ \\
		$\mathrm{L}656\_\mathrm{M}11\_\mathrm{frcdm}$ & $f(R)$CDM & $656.25$ & $512^{3}$ & $1.79\cdot10^{11}$ & $8$ & $80$ \\
        \hline
		$\mathrm{L}328\_\mathrm{M}9\_\mathrm{wcdm}$ & $w$CDM & $328.125$ & $1024^{3}$ & $2.81\cdot10^{9}$ & $1$ & $16$ \\
		$\mathrm{L}656\_\mathrm{M}10\_\mathrm{wcdm}$ & $w$CDM & $656.25$ & $1024^{3}$ & $2.25\cdot10^{10}$ & $1$ & $16$ \\
        \hline
		$\mathrm{L}328\_\mathrm{M}10\_\mathrm{P}18$ & P$18$ & $328.125$ & $512^{3}$ & $2.25\cdot10^{10}$ & $384$ & $1$ \\
		$\mathrm{L}328\_\mathrm{M}10\_\mathrm{w}12$ & w$12$ & $328.125$ & $512^{3}$ & $2.25\cdot10^{10}$ & $64$ & $1$ \\
		$\mathrm{L}328\_\mathrm{M}10\_\mathrm{F}5$ & F$5$ & $328.125$ & $512^{3}$ & $2.25\cdot10^{10}$ & $64$ & $1$ \\
		\hline
        $\mathrm{L}164\_\mathrm{M}10$ & P$18$, F$5$, F$6$ & $164.0625$ & $256^{3}$ & $2.25\cdot10^{10}$ & $1$ & $3$ \\
        $\mathrm{L}164\_\mathrm{M}9$ & P$18$, F$5$, F$6$ & $164.0625$ & $512^{3}$ & $2.81\cdot10^{9}$ & $1$ & $3$ \\
        $\mathrm{L}164\_\mathrm{M}8$ & P$18$ & $164.0625$ & $1024^{3}$ & $3.51\cdot10^{8}$ & $1$ & $1$ \\
        \hline
	\end{tabular}
    \tablefoot{
      The first column correspond to the identifier of the simulation suite (Name), the second column indicates the cosmological scenario of the simulation suite (Model), the third column shows the length of the simulation box ($L_{\mathrm{box}}$), the fourth column the number of $N$-body particles ($N_{\mathrm{part}}$), the fifth column gives the associated $N$-body particle mass ($m_{\mathrm{part}}$) in units of $\left(\Omega_{\rm m}/0.3071\right)h^{-1}M_{\odot}$, the sixth column gives the number of independent realization per model ($N_{\mathrm{real}}$), while the last column gives the number of cosmological models simulated for each cosmological scenario ($N_{\mathrm{cosmo}}$).
    }
\end{table*}

The extended \textsc{e-mantis} simulation suite consists of simulations of different cosmological models that pave the parameter space of the $w$CDM and $f(R)$CDM models.
For each of the sampled models, the suite includes simulations of different volume, mass resolution and number of independent realizations.
In this subsection, we describe the main characteristics of these simulations, which are summarized in Table~\ref{table:sim_suite}.

The core set of simulations, which are used to train the emulators, sample $80$ cosmological models using the strategy described in Sect.~\ref{subsec:param_space_sampling}, both in $w$CDM and $f(R)$CDM.
For each cosmological model, we use two different simulation volumes, a smaller box of side length $L_\mathrm{box} = 328.125\,h^{-1}\mathrm{Mpc}$ (hereafter L$328$) and a larger box of side length $L_\mathrm{box}= 656.25\,h^{-1}\mathrm{Mpc}$ (hereafter L$656$).
In both cases we use $512^{3}$ $N$-body particles, resulting in a particle mass resolution of $m_{\rm part}\sim 10^{10}\,h^{-1}M_{\odot}$ (hereafter M$10$) and $m_{\rm part}\sim 10^{11}\,h^{-1}M_{\odot}$ (hereafter M$11$) respectively.
Note that the exact particle mass depends on the value of $\Omega_{\rm m}$  used in the simulation (see Table~\ref{table:sim_suite}).
The resolution of the $\mathrm{L}328\_\mathrm{M}10$ simulations is sufficient to resolve group sized haloes (see Sect.~\ref{subsec:mass_res_corr} and appendix~\ref{app:numerical_convergence}), while the volume of the $\mathrm{L}656\_\mathrm{M}11$ set provides sufficient statistics to study the abundance of cluster sized haloes.
The systematic error on the high mass end of HMF due to the finite volume of the L$656$ box is smaller than $\sim2\%$ up to halo masses of $M_h\lesssim10^{15}\,h^{-1}M_\odot$ \citep{Castro_2023}.
This is negligible with respect to the Poisson noise (see Fig.~\ref{fig:hmf_bspline_fit}).
For each cosmological model and each box-size, we run $64$ independent realizations in the case of $w$CDM and $8$ in the case of $f(R)$CDM, as the computation is much more intensive in the latter case. 
We thus reach an effective volume of $\left(2625\,h^{-1}\mathrm{Mpc}\right)^3$ per cosmological model in $w$CDM, and $\left(1312.5\,h^{-1}\mathrm{Mpc}\right)^3$ in $f(R)$CDM.
The initial conditions are common to all cosmological models.
The random phases of the initial conditions have been designed such that each group of $8$ realizations of the $\mathrm{L}328\_\mathrm{M}10$ simulations correspond to the same random phase as one of the $\mathrm{L}656\_\mathrm{M}11$ simulations.
This means that the first $\mathrm{L}656\_\mathrm{M}11$ realizations bring no additional information with respect to the whole set of $\mathrm{L}328\_\mathrm{M}10$ simulations.
However, we can use this matching to correct for finite mass resolution errors in the $\mathrm{L}656\_\mathrm{M}11$ simulations, and for each individual cosmological model.

In addition to this core set, we run higher resolution simulations with $1024^{3}$ $N$-body particles for both the $\mathrm{L}328$ and $\mathrm{L}656$ box-sizes, which correspond to a particle mass resolution of $m_{\rm part}\sim 10^{9}\,h^{-1}M_{\odot}$ (hereafter M$9$) and $m_{\rm part}\sim 10^{10}\,h^{-1}M_{\odot}$ (M$10$) respectively.
We use these simulations to correct for finite mass resolution effects.
We run these simulations only for a single realization and only for the first slice of our $w$CDM quasi-PLHS, which corresponds to a perfect LHS with $16$ models.
Therefore, it is possible to partially explore the cosmological dependence of such finite mass resolution effects in $w$CDM.

We run additional simulations for some reference cosmological models.
In particular, we consider a \citet{Planck2018} compatible $\Lambda$CDM cosmology (hereafter $\mathrm{P}18$), with $\Omega_{\mathrm{m}} = 0.3071$, $\sigma_{8}^{\rm GR} = 0.8224$, $h = 0.6803$, $n_{\mathrm{s}} = 0.96641$ and $\Omega_{\mathrm{b}} = 0.048446$.
We also consider a reference $w$CDM cosmology with $w=-1.2$ (hereafter $\mathrm{w}12$) and a reference $f(R)$CDM model with $f_{R_{0}} = -10^{-5}$ (hereafter $\mathrm{F}5$).
In both cases, the other cosmological parameters take the values of the P$18$ model.
The values of these cosmological parameters are summarized in Table~\ref{table:cosmo_params}.
We run $384$ independent realizations of the $\mathrm{L}328\_\mathrm{M}10$ box for P$18$ and $64$ realizations for w$12$ and F$5$.
These simulations are not used during the emulator training and serve only as a validation set.

Finally, we run another set of simulations in order to study the numerical convergence in terms of mass resolution for some selected reference models.
This last set uses a box-size of $L_\mathrm{box}=164.0625\,h^{-1}\mathrm{Mpc}$ (hereafter L$164$), with $256^{3}$, $512^{3}$ and $1024^{3}$ $N$-body particles, corresponding to a particle mass resolution of $m_{\rm part}\sim 10^{10}\,h^{-1}M_{\odot}$ (M$10$), $m_{\rm part}\sim 10^{9}\,h^{-1}M_{\odot}$ (M$9$) and $m_{\rm part}\sim 10^{8}\,h^{-1}M_{\odot}$ (hereafter M$8$) respectively.
We run a single realization for the P$18$ cosmology for each of these resolutions, and for the F$5$ and F$6$ cosmologies for the M$10$ and M$9$ resolutions.

We store all the particles of the simulation snapshots, as well as the particles of FoF and SO haloes at $11$ redshifts: $z=\{z_{i}\sim49, 2.6, 2, 1.5, 1.25, 1, 0.7, 0.5, 0.25, 0.1, 0\}$.
We save the particle positions and velocities, as well as the values of the two Bardeen potentials ($\phi$ and $\psi$) and the value of the $f_{R}$ field interpolated at the position of each particle.
Additionally, we store the matter power spectra and halo profiles, as well as some other basic halo properties, for both FoF and SO haloes at $23$ redshift outputs: $z=\{z_{i}\sim49, 3, 2.6, 2.3, 2, 1.74, 1.5, 1.4, 1.25, 1.1$, $1, 0.9, 0.7, 0.6, 0.5, 0.42, 0.3, 0.25, 0.15, 0.1, 0.05, 0\}$.

\section{Emulating the HMF}
\label{subsec:emulating_hmf}

The halo mass function quantifies the number of haloes of a given mass in an infinitesimal comoving volume.
The seminal work by \citet{PS_hmf} provided the first analytical derivation of the HMF within the hierarchical bottom-up CDM scenario of cosmic structure formation.
Subsequently, \citet{Bond1991} formulated a stochastic model to predict the HMF in a given cosmological scenario.
However, the discrepancy between the analytical predictions and the results of simulations has led to the development of a phenomenological approach driven by fitting parameterized functions to the results of $N$-body simulations.
Building upon the work of \citet{ST1999}, over the last two decades numerous works in the literature have presented calibrated parametrization of the HMF improving on the resolution and volume of $N$-body simulations with the intent of better capturing its mass, redshift, and cosmology dependence. 

The advent of a new generation of galaxy cluster surveys has triggered novel effort in the development of fast and accurate predictions of the HMF.
The realization of emulators capable of providing predictions accurate at the level required to perform unbiased cosmological analyses for a wide range of cosmological parameter values and different cosmological scenarios is therefore timely.
Here, we will present the simulation training data and the HMF emulation algorithm implemented in the \textsc{e-mantis} emulator. 

\subsection{The HMF training data}
\label{subsec:hmf_training_data}

We detect FoF dark matter haloes with a linking length of $b=0.2$ (in units of the mean interparticle distance), and SO haloes with an overdensity threshold of $\Delta=200$ times the critical density of the universe (hereafter $200\mathrm{c}$).
From the density profiles of the SO haloes detected with $\Delta=200\mathrm{c}$, we compute for each halo their mass at the overdensity thresholds $\Delta=500\mathrm{c}$ and $\Delta=1000\mathrm{c}$.
We use these additional conditional mass measurements as new halo definitions.
We estimate the halo mass function, $\mathrm{d}n/\mathrm{dln}M_{h}$, for all these halo definitions, with constant logarithmic mass bins of width $\Delta\log_{10}{M_{h}}\simeq0.05$.
Halo masses are given in units of $h^{-1}M_{\odot}$, thus implying that the HMF has units of $h^{3}\,\mathrm{Mpc}^{-3}$, where $\mathrm{Mpc}$ denotes comoving megaparsec.

The HMF emulator training data is built by combining the HMF measurements from the simulation boxes $\mathrm{L}328\_\mathrm{M}10$ and $\mathrm{L}656\_\mathrm{M}11$ of the emulator set (see Sect.~\ref{subsec:nbody_suite}).
More specifically, we consider a minimum mass in the small volume simulations of $M_{h, \mathrm{min}} = 5\cdot10^{12} \, h^{-1}M_{\odot}$ in the case of FoF haloes, while in the case of the SO haloes we set $M_{h, \mathrm{min}} = 10^{13} \, h^{-1}M_{\odot}$.
These values correspond to haloes with $\sim150$ and $\sim300$ particles, respectively, for the models with the worst mass resolution (corresponding to cosmologies with $\Omega_\mathrm{m} = 0.465$).
As shown in Sect.~\ref{subsec:mass_res_corr}, this is the minimum number of particles for which it is possible to apply a cosmology independent mass resolution correction.
We use the $\mathrm{L}328\_\mathrm{M}10$ HMF up to a mass corresponding to haloes with $N_\mathrm{part, M11}=2000$ low resolution particles (i.e. $N_{\rm part, M10}=16000$ high resolution particles), and switch to the $\mathrm{L}656\_\mathrm{M}11$ HMF for haloes with a larger number of particles.
We find that this threshold ensures a smooth transition from one simulation box to the other.
Any residual systematic difference is compensated by the resolution correction introduced in Sect.~\ref{subsec:mass_res_corr}.
We force a minimum number of $5$ haloes per bin by merging high mass end bins.
For each cosmological model $\theta_{i}$, we halt the computation of the HMF at a maximum mass $M_{h,\mathrm{max},i}$, corresponding to the mass bin in which the criterion can no longer be satisfied.
We estimate the errors on the HMFs by using $1000$ bootstrap random catalogues.

\subsection{HMF mass resolution correction}
\label{subsec:mass_res_corr}

\begin{figure*}
	\centering
	\begin{subfigure}{0.49\linewidth}
		\includegraphics[width=0.95\linewidth]{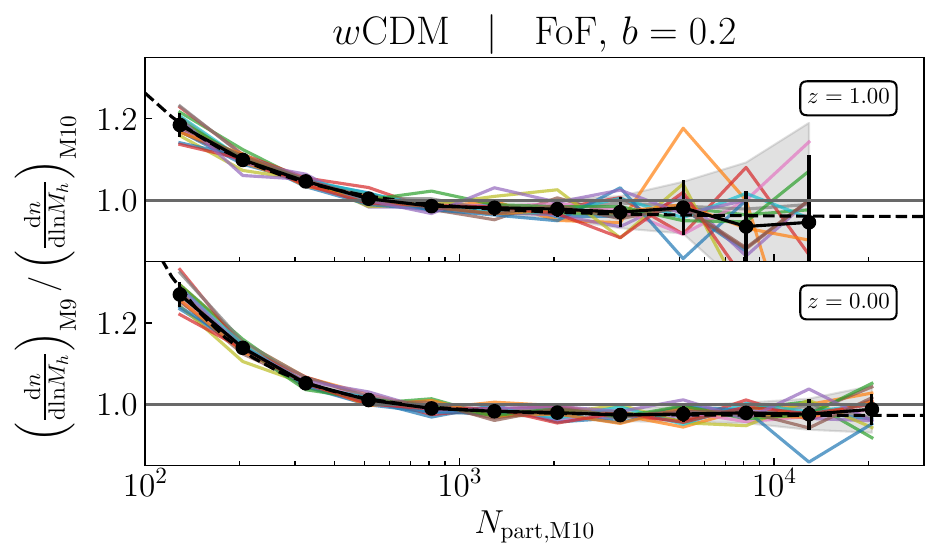}
	\end{subfigure}
	\begin{subfigure}{0.49\linewidth}
		\includegraphics[width=0.95\linewidth]{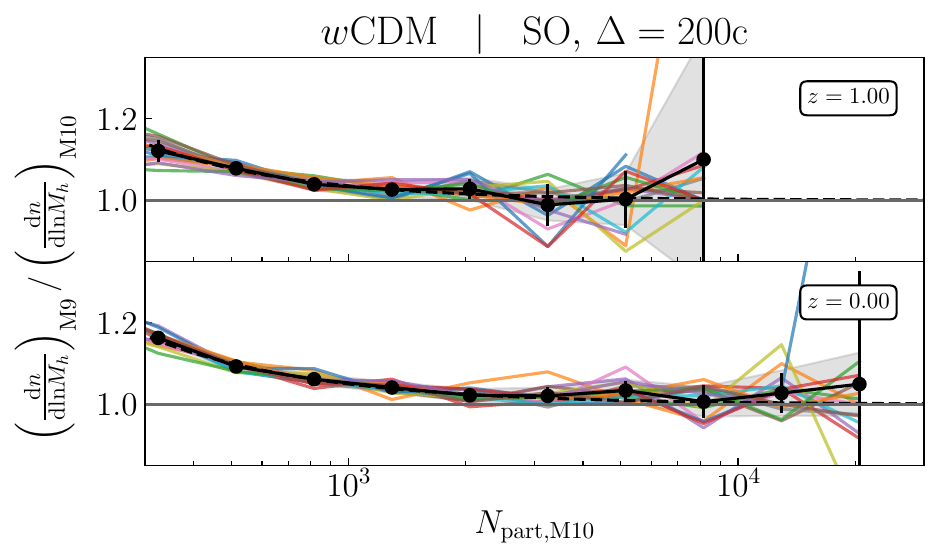}
	\end{subfigure}
	\begin{subfigure}{0.49\linewidth}
		\includegraphics[width=0.95\linewidth]{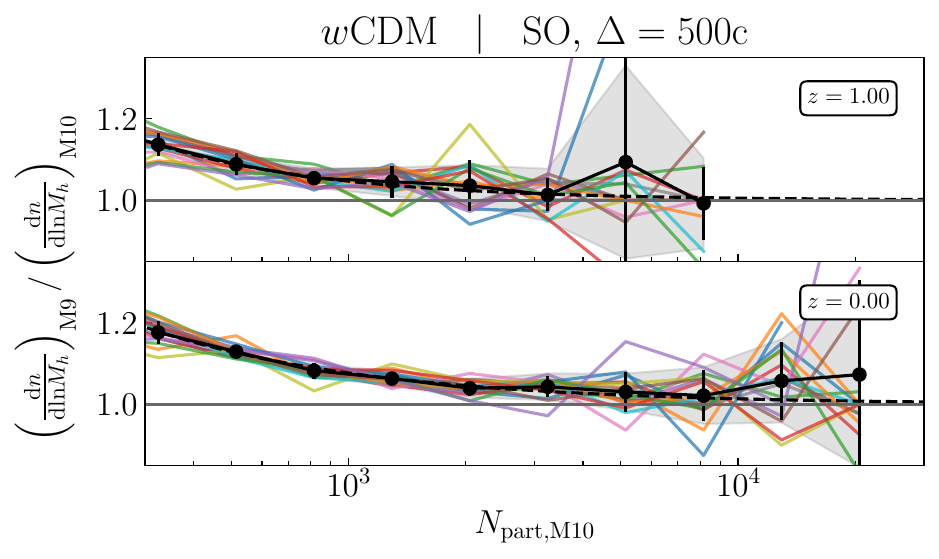}
	\end{subfigure}
	\begin{subfigure}{0.49\linewidth}
		\includegraphics[width=0.95\linewidth]{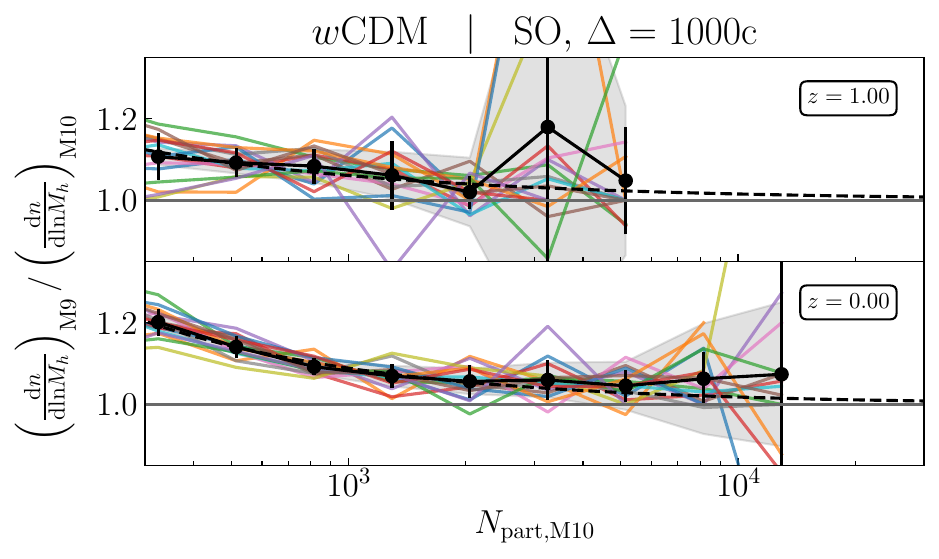}
	\end{subfigure}
	\caption{
      Ratio of the HMF with the better mass resolution M$9$ to the one with the standard mass resolution M$10$ as a function of the number of standard resolution particles per halo $N_{\rm part, M10}$ at $z=0$ and $z=1$.
      This quantity has been computed using the higher resolution simulation set $\mathrm{L}328\_\mathrm{M}9$.
      Each sub-plot shows the case of a different halo definition: FoF with $b=0.2$ (\textit{top-left}) and SO with $\Delta=200{\rm c}$ (\textit{top-right}), $\Delta=500{\rm c}$ (\textit{bottom-left}) and $\Delta=1000{\rm c}$ (\textit{bottom-right}).
      Each coloured line corresponds to one of the $16$ cosmological models of the first slice of the $w$CDM quasi-PLHS.
      The black solid line gives the mean over all cosmological models, with the standard deviation among them as error bars.
      \corr{The grey band attached to the mean is an estimation of the typical expected error for the individual HMF ratio.}
      The dashed line corresponds to the fit to the mean using the fitting function given in Eq.~(\ref{eq:mass_corr_SO}) for SO haloes, and in Eq.~(\ref{eq:mass_corr_FOF}) for the FoF case.
	}
	\label{fig:mass_res_corr_M9_M10}
\end{figure*}

\begin{figure*}
  \begin{subfigure}{0.49\linewidth}
	\includegraphics[width=0.95\linewidth]{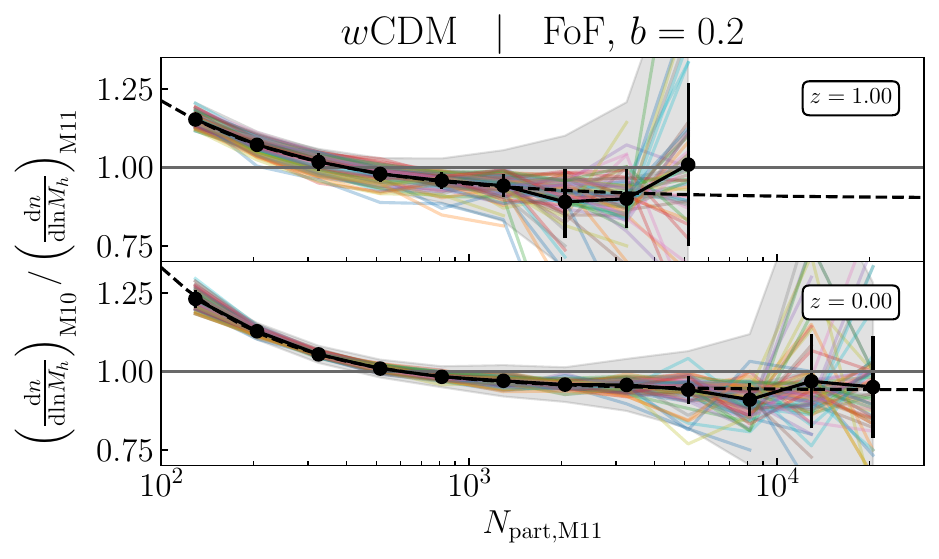}
  \end{subfigure}
  \begin{subfigure}{0.49\linewidth}
	\includegraphics[width=0.95\linewidth]{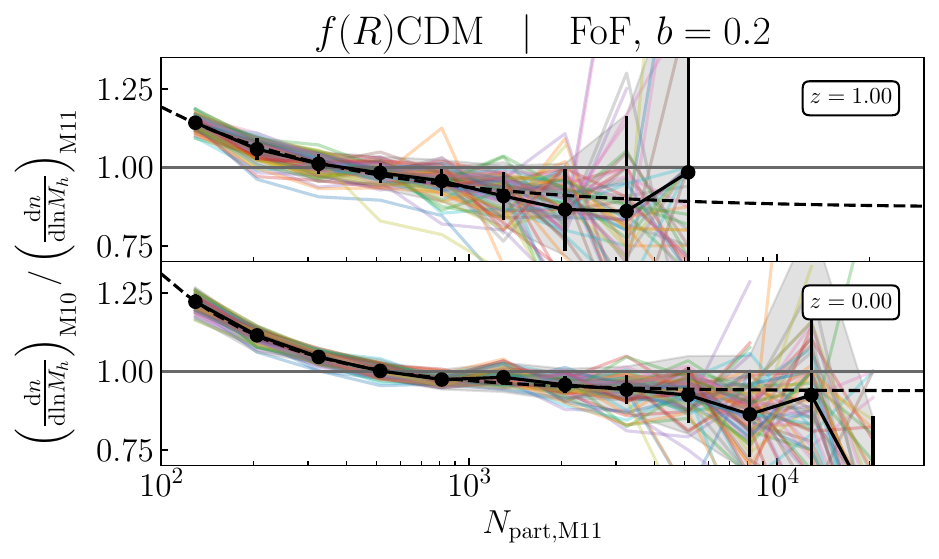}
  \end{subfigure}
  \caption{
      Ratio of the FoF ($b=0.2$) HMF with the better mass resolution M$10$ to the one with the standard mass resolution M$11$ as a function of the number of standard resolution particles per halo $N_{\rm part, M11}$, in $w$CDM (\textit{left}) and $f(R)$CDM (\textit{right}), at $z=0$ and $z=1$.
      This quantity has been computed using the matching in terms of initial conditions between the $\mathrm{L}656\_\mathrm{M}11$ and the $\mathrm{L}328\_\mathrm{M}10$ simulation sets.
      Each coloured line corresponds to one of the $80$ cosmological models of the emulator training set.
      The black solid line gives the mean over all cosmological models, with the standard deviation among them as error bars.
      \corr{The grey band attached to the mean is an estimation of the typical expected error for the individual HMF ratio.}
      The dashed line corresponds to the fit to the mean using the fitting function given in Eq.~(\ref{eq:mass_corr_FOF}).
	}
  \label{fig:mass_res_corr_M10_M11}
\end{figure*}

In order to both improve the accuracy of the $\mathrm{L}328\_\mathrm{M}10$ training data at the low mass end, and to smooth out the transition between the $\mathrm{L}656\_\mathrm{M}11$ and the $\mathrm{L}328\_\mathrm{M}10$ simulation boxes, we compute a correction for the finite mass resolution errors in the HMF measurements.

We define the correction as the ratio of the HMF measured from a simulation with a higher resolution $\mathrm{M}X$ (i.e. a lower particle mass) to the HMF measured from a lower resolution $\mathrm{M}Y$ (i.e. a higher particle mass), with $X<Y$.
In practice, we do not dispose of high resolution simulations for all the cosmological models and realizations of our simulation suite.
As such, the first approximation we employ, is to assume that the correction will be independent of the specific realization.
Additionally, we find that when the correction is expressed in terms of the low resolution number of particles per halo $N_{\mathrm{part, M}Y}$, instead of the halo mass $M_{h}$, the correction is independent of the cosmological model.
This approximation holds up to a certain minimum number of particles per halo.
A direct consequence is that we can average the HMF ratio over multiple cosmological models in order to reduce the measurement noise.
It is also possible to apply the correction computed on a subset of cosmological models to all the training data of the emulator simulation set.
Furthermore, in order to obtain a smooth correction, we fit an analytical function to the measured HMF ratio averaged over all available cosmological models and realizations.
In the case of SO haloes, we use the following fitting function for the ratio of the HMF with the higher resolution $\mathrm{M}X$ and the HMF with the lower resolution $\mathrm{M}Y$
\begin{equation}
  \label{eq:mass_corr_SO}
  \frac{\left(\frac{\mathrm{d}n}{\mathrm{dln}M_h}\right)_{\mathrm{M}X}}{\left(\frac{\mathrm{d}n}{\mathrm{dln}M_h}\right)_{\mathrm{M}Y}} = 1 + \exp \left(-\log_{10}\left(N_{\mathrm{part, M}Y}/N_{0}\right)/\sigma_{0}\right),
\end{equation}
where $N_{0}$ and $\sigma_{0}$ are free dimensionless parameters.
In the case of FoF haloes, we modify the fitting function as
\begin{equation}
  \label{eq:mass_corr_FOF}
  \frac{\left(\frac{\mathrm{d}n}{\mathrm{dln}M_h}\right)_{\mathrm{M}X}}{\left(\frac{\mathrm{d}n}{\mathrm{dln}M_h}\right)_{\mathrm{M}Y}} = c + \exp \left(-\log_{10}\left(N_{\mathrm{part, M}Y}/N_{0}\right)/\sigma_{0}\right),
\end{equation}
where $c$ is an additional free dimensionless parameter.

In order to improve the accuracy of the HMF training data at the low mass end, we compute this type of correction for the $\mathrm{L}328\_\mathrm{M}10$ simulation boxes of the emulator training set.
To do so, we measure the corresponding HMF ratio using the $\mathrm{L}328\_\mathrm{M}9\_\mathrm{wcdm}$ simulation set described in Sect.~\ref{subsec:nbody_suite}.
Figure~\ref{fig:mass_res_corr_M9_M10} shows the HMF ratios for the $16$ cosmological models of the first $w$CDM quasi-PLHS, along with their mean and standard deviation, for different halo definition and redshifts.
\corr{
  We also compute the expected error on each HMF ratio using $4^{3}$ spatial jackknife sub-volumes.
  The different cosmological models have different errors.
  For simplicity, we average all the errors in quadrature and show them in Fig.~\ref{fig:mass_res_corr_M9_M10} attached to the mean HMF ratio.
  This mean error is indicative of the typical expected statistical noise in each HMF ratio.
  Its agreement with the standard deviation between the different models is indicative of the cosmological independece of the mass resolution correction.
}

We find that the mass resolution correction is cosmology independent for FoF haloes with $b=0.2$ and $N_{\rm part, M10} \gtrsim 150$ and SO haloes with $\Delta=200\mathrm{c}$, $\Delta=500\mathrm{c}$ and $N_{\rm part, M10} \gtrsim 300$.
For a smaller number of particles, we find large deviations from one cosmology to another (not shown in Fig.~\ref{fig:mass_res_corr_M9_M10}).
Since we want to average the HMF ratio over all cosmologies in order to reduce the noise in the signal, this observation motivates the minimum mass supported by the emulator (see Sect.~\ref{subsec:hmf_training_data}).
We stress the fact that the $16$ cosmological models considered form a perfect LHS, and therefore explore the whole $6$-dimensional $w$CDM parameter space, albeit with a reduced number of points.
In the case of SO haloes with $\Delta=1000\mathrm{c}$, the measurements are noisier, which makes it more difficult to conclude on the cosmological dependence.
However, considering a correction beyond the mean cosmological HMF ratio would require larger volume simulations or more realizations of the high resolution $\mathrm{L}328\_\mathrm{M}9$ box.
We decide to keep the same minimum number of particles per halo as for the other overdensities, i.e. $N_{\mathrm{part,M}10}\gtrsim300$.

In the case of SO haloes, the HMF ratio has the shape of a decreasing exponential, converging towards unity at large number of particles per halo.
\corr{
  This feature is mainly due to the dynamical solver of the $N$-body code, which, because of the finite resolution, tends to underestimate the clustering at small scales} \citep[see e.g.][]{Rasera_2014}.
In the case of FoF haloes, the HMF ratio does not converge to unity even at large number of particles per halo.
There is a residual shift of a few percent.
\corr{
  Indeed, FoF haloes are affected by an additional resolution effect.
  As pointed out by \citet{More_2011}, the FoF algorithm tends to overestimate the mass of a halo if the resolution of the simulation is not sufficient (i.e. if the halo is not sampled by enough particles).
  Our interpretation is that, for small haloes, the error from the dynamical solver dominates.
  When this component converges to unity, the opposite error due to the FoF algorithm has not yet converged.
  This makes the HMF ratio crossing the unity line.
  Eventually, the HMF ratio should converge to one.
  However, the convergence is slow and our data is noisy at the high mass end.
  We decide to approximate this regime by a constant, corresponding to the extra parameter in the fitting formulan given in Eq.~(\ref{eq:mass_corr_FOF}) with respect to the SO case.
}
\corr{
  In practice, we never use the correction for haloes larger than the last data point for which we are able to measure the HMF ratio, since the mass range of the emulator does not go beyond that.
}

We fit the analytical forms of Eqs.~(\ref{eq:mass_corr_SO}) and~(\ref{eq:mass_corr_FOF}) to the mean HMF ratio, independently for each halo definition and redshift.
The resulting correction, which is also shown in Fig.~\ref{fig:mass_res_corr_M9_M10}, is applied to all the HMF training data of both $w$CDM and $f(R)$CDM emulators.
In App.~\ref{app:numerical_convergence}, we show that even if this correction has been computed from $w$CDM simulations, it remains valid in the case of $f(R)$ gravity.
We also test the numerical convergence of the correction.

Although the transition between the simulation boxes $\mathrm{L}328\_\mathrm{M}10$ and $\mathrm{L}656\_\mathrm{M}11$ takes place at a high number of low resolution number of particles per halo (see Sect.~\ref{subsec:hmf_training_data}), we find that in some cases a small residual systematic difference remains.
This produces a discontinuity in our HMF training data, which degrades the accuracy of the final emulator.
In principle, we could use the $\mathrm{L}656\_\mathrm{M}10\_\mathrm{wcdm}$ simulation set, to compute the same kind of correction as described in the last paragraph.
However, in this case the HMF ratio is significantly noisier, since we are interested in more massive haloes.
The larger volume of the $\mathrm{L}656$ simulation box with respect to the $\mathrm{L}328$ box does not compensate the drop of the HMF.
Therefore, even by averaging the HMF ratio over the $16$ cosmological models of the $\mathrm{L}656\_\mathrm{M}10\_\mathrm{wcdm}$ set, the measurements are too noisy to perform a proper fit at all redshifts.
Instead, we make use of the initial condition matching between the $\mathrm{L}328\_\mathrm{M}10$ and the $\mathrm{L}656\_\mathrm{M}11$ simulation sets described in Sect.~\ref{subsec:nbody_suite}.
In this case, we can compute the mean correction using $80$ cosmological models instead of $16$, and therefore obtain a smoother measurement.
We compute the ratio of the HMF measured using all the realizations from $\mathrm{L}328\_\mathrm{M}10$ and the HMF from the $\mathrm{L}656\_\mathrm{M}11$ realizations corresponding to the same initial random phase, both in $w$CDM and $f(R)$CDM.
We perform the same kind of fit to the mean HMF ratio and apply the obtained correction to all the measurements from the $\mathrm{L}656\_\mathrm{M}11$ simulation box.
We find that, in the case of SO haloes, at the transition between both simulations volumes, i.e. for a number of low resolution particles per halo of $N_{\rm part, M11} = 2000$, the $\mathrm{L}656\_\mathrm{M}11$ HMF has converged to the $\mathrm{L}328\_\mathrm{M}10$ HMF within the statistical noise of the training data.
Therefore, the transition between both simulation volumes is smooth without the need to correct the resolution of the larger volume simulations.
This is due to the fact that we have chosen to only use the $\mathrm{L}656\_\mathrm{M}11$ data for sufficiently massive haloes.
We still apply the correction, both for consistency and to remove any residual errors.
However, in the case of FoF haloes, we observe the same kind of systematic shift at large number of particles per halo as discussed in the previous paragraph.
Therefore, it is essential to correct this effect to properly transition from one simulation volume to the other.
We find that if this correction is not applied, the goodness-of-fit of the HMF B-spline decomposition presented in Sect.~\ref{subsec:bspline_fit} is degraded significantly around the transition, i.e. $N_{\rm part, M11}\simeq 2000$.
The HMF ratios and the corresponding mean and fitted function, for the FoF case both in $w$CDM and $f(R)$CDM, are shown in Fig.~\ref{fig:mass_res_corr_M10_M11}.

\subsection{B-spline decomposition of the HMF}
\label{subsec:bspline_fit}

\begin{figure}
	\centering
	\includegraphics[width=0.95\linewidth]{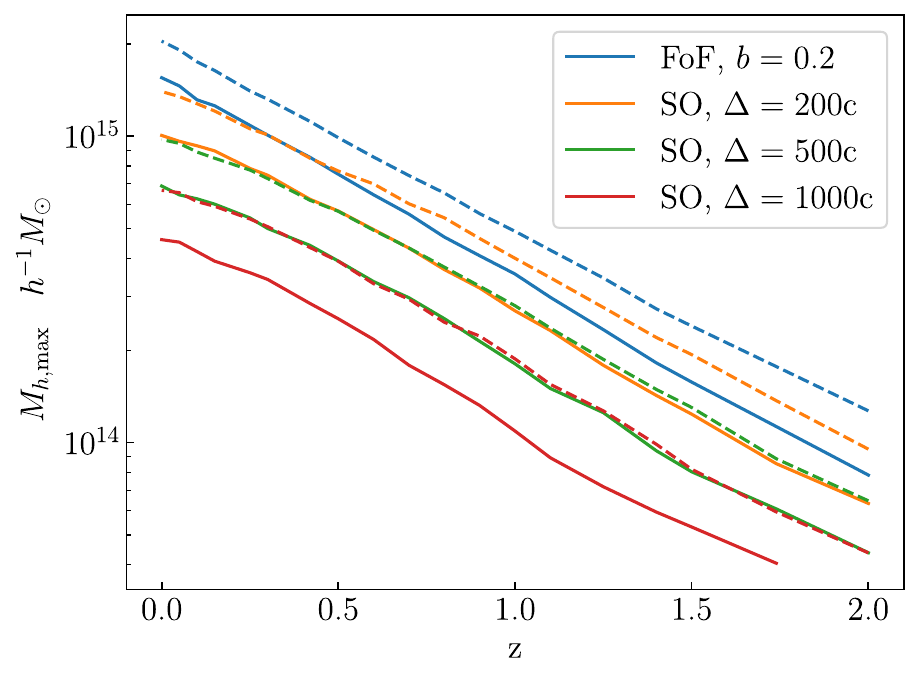}
	\caption{
		Maximum halo mass range of the $f(R)$ emulator (solid lines) and $w$CDM emulator (dashed lines) as a function of redshift and for different halo definitions, as defined by Eq.~(\ref{eq:mass_range}).
	}
	\label{fig:emu_mass_range}
\end{figure}

One common approach to emulate the HMF is to first fit the measured HMFs of each cosmological model using some parametric fitting formula.
In a second step, the coefficients from the fitted HMFs are emulated to account for their cosmological dependence.
As an example, a number of $w$CDM emulators of the HMF have assumed the parametrization from \citet{Tinker_2008} as fitting function \citep{aemulus_hmf_emu, dark_quest_hmf_emu}.
However, previous studies have shown that this type of approach is not able to accurately capture the shape of the HMF in modified gravity theories at percent level accuracy \citep[see e.g.][]{schmidt_fR_hmf, lombriser_fR_hmf, Gupta_2022}.
Because of this, we choose a different approach and adopt a fitting strategy that limits the assumptions made about the shape of the HMF.
\corr{
  We fit the numerically estimated HMF by piecewise polynomials or splines, following a similar procedure to the one used in \citet{mira_titan_hmf_emu}.
  More specifically, we decompose the HMF using B-splines, which constitute a basis for spline functions, while the work of \citet{mira_titan_hmf_emu} uses constrained piecewise polynomials.
}

Formally, a B-spline decomposition is defined by the degree $p$ of the \corr{polynomial pieces}, the positions $t_k$ of the knots (i.e. where the polynomial pieces meet) and the coefficients $c_j$ multiplying each B-spline basis function.
\corr{
  Consider two boundary knots delimiting the range of the B-spline decomposition and an ensemble of $n$ internal knots, which corresponds to $n+1$ polynomial pieces.
  Constructing the B-spline basis requires $p$ additional knots below the lower boundary knot and another $p$ knots above the upper one \citep[see e.g.][]{Perperoglou_2019}.
  Their values are arbitratry, and they are usually set equal to the lower and upper boundaries, respectively.
  Once the knots have been fixed, the B-splines of degree $p=0$ are defined as
}
\begin{equation}
	B_{j,0}\left[x\right] \equiv \left\{\begin{array}{@{}l@{}}
		1 \quad \mathrm{if} \, t_j \corr{\leq} x < t_{j+1}, \\
		0 \quad \mathrm{otherwise}.
	\end{array}\right.
\end{equation}
Then, B-splines of degree $p>0$ can be constructed by recursion \corr{\citep{Deboor_1972}}
\begin{equation}
	B_{j,p}\left[x\right] \equiv \frac{x-t_j}{t_{j+p}-t_j}B_{j, p-1;\mathbf{t}}\left[x\right] + \frac{t_{j+p+1}-x}{t_{j+p+1}-t_{j+1}}B_{j+1, p-1;\mathbf{t}}\left[x\right],
\end{equation}
\corr{for $j=1,\cdots,n+p+1$.}
The complete B-spline basis is made up of $\corr{n+p+1}$ functions.
Hence, the B-spline decomposition of the estimated HMF reads as
\begin{equation}
	\log_{10} \left[\frac{\mathrm{d}n}{\mathrm{dln}M_h}\right] = \sum_{\corr{j=1}}^{\corr{n+p+1}}c_jB_{j,p}\left[\log_{10}M_h\right].
\end{equation}
We choose splines of degree $p=2$ and knots of  logarithmic mass points equally spaced by $\Delta\log_{10} M_{h} \simeq 0.25$.
The only assumption about the shape of the HMF is that it can be described by a polynomial of degree $2$ in log-log axes over intervals of $\sim0.25$ dex.
The same description of the shape of the $w$CDM HMF is used in~\citet{mira_titan_hmf_emu}.
We find that this approach is also able to capture the shape of the HMF in $f(R)$ gravity.

\begin{figure*}
	\centering
	\begin{subfigure}{0.49\linewidth}
		\includegraphics[width=0.95\linewidth]{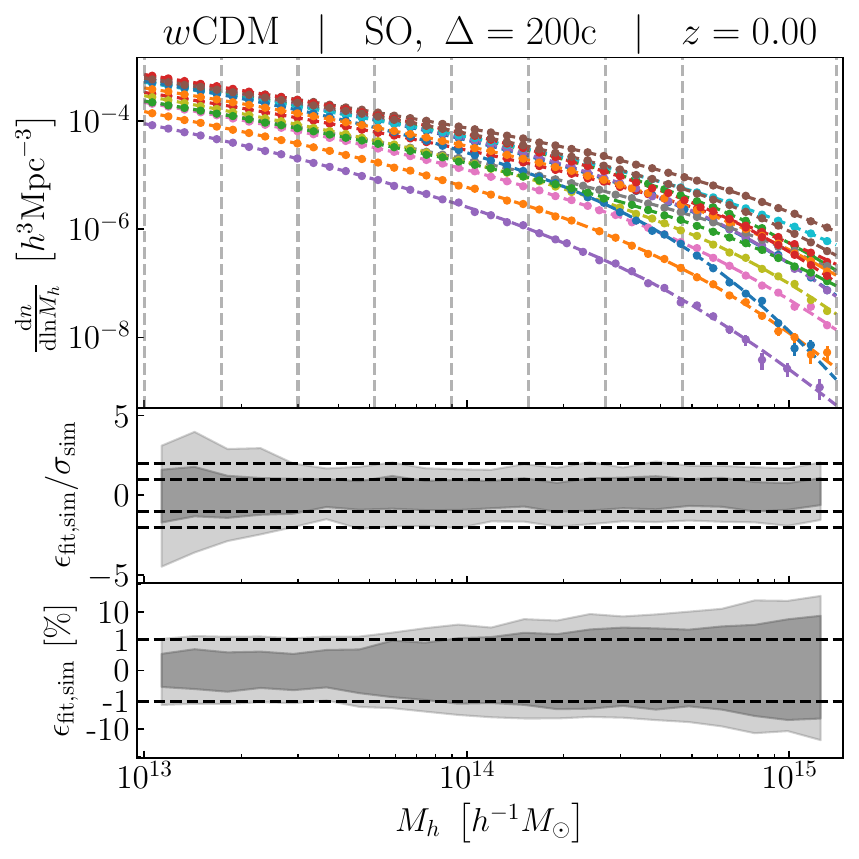}
	\end{subfigure}
	\begin{subfigure}{0.49\linewidth}
		\includegraphics[width=0.95\linewidth]{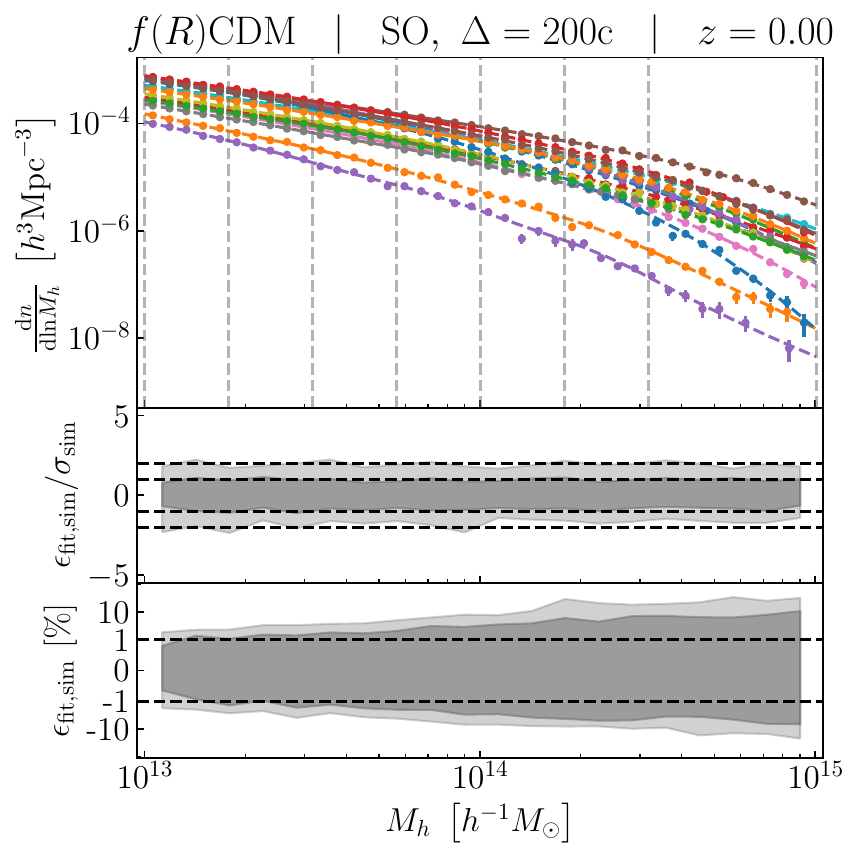}
	\end{subfigure}
	\caption{
		B-spline decomposition of the HMF for SO haloes with $\Delta=200\mathrm{c}$ at $z=0$ in $w$CDM (\textit{left}) and $f(R)$CDM (\textit{right}) cosmologies.
		Top panel: measured HMF from the simulations (data points), with the corresponding errors bars (not always visible), and B-spline function fits (dashed lines).
		Each colour corresponds to a different training model.
        For visual clarity, only the models from the first slice of the quasi-PLHS are displayed (i.e. $16$ models).
        The dashed vertical lines indicate the location of the B-spline knots.
		Middle panel: distribution of the fit residuals normalised by the noise in the simulation data.
        The dark and light shaded areas mark the $68.2\%$ and $95.4\%$ confidence intervals, respectively.
        The dashed lines indicate the $\pm1$ and $\pm2\sigma$ levels.
		Bottom panel: distribution of the fit residuals normalized by the simulation data (i.e. the relative errors).
        The dark and light shaded areas mark the $68.2\%$ and $95.4\%$ confidence intervals, respectively.
		The shaded region corresponds to the $\pm1\%$ level, along the y-axis the scaling is linear within the $1\%$ region and logarithmic outside.
		The uncertainties on the B-spline fits are dominated by the simulation noise, except in the $w$CDM case at small masses, where they saturate at the $~\sim0.5\%$ level.
	}
	\label{fig:hmf_bspline_fit}
\end{figure*}

One important aspect of the B-spline decomposition to be considered, is that when fitting the HMF, the maximum mass needs to be the same for all cosmological models, since we want to decompose all the individual HMFs into the same B-spline basis.
However, we find that there is up to an order of magnitude difference between the maximum mass of different cosmological models.
We set the maximum mass \corr{of the B-spline decomposition} to
\begin{equation}
\label{eq:mass_range}
 \corr{M_{h, \mathrm{max}} = \alpha\times10^{\left<\log_{10 }M_{h,\mathrm{max},i}\right>_{\theta_{i}}}},
\end{equation}
where the average is performed over all cosmological models $\theta_{i}$.
We find that a value of $\alpha=0.5$ gives a good compromise between maximizing the range of the emulator without degrading the quality of the B-spline fit at the high-mass end. 
This defines the maximum mass range of the emulator at each redshift and for each halo definition, as shown in Fig.~\ref{fig:emu_mass_range}.
Furthermore, it is necessary that for each cosmological model, there is a minimum number of data points per polynomial piece to perform the B-spline decomposition.
For this purpose, we merge the high mass \corr{internal} knots till all the cosmological models have at least three data points in the last spline segment.
In Fig.~\ref{fig:hmf_bspline_fit} we show a visual example of the location of the B-spline knots.

We use the Python package \textsc{scipy}~\citep{scipy} to perform the spline fitting to the HMFs and include the diagonal errors of the simulation measurements into the fit.
We estimate the variance of the spline coefficients by drawing $1000$ random HMFs from the estimated simulation noise, and repeating the fitting procedure each time.
We repeat this operation for each halo definition and each redshift output of the $w$CDM and $f(R)$CDM simulation sets. In Fig.~\ref{fig:hmf_bspline_fit} we compare the B-spline fit to the HMF data measured from the simulations for the case of SO haloes with $\Delta=200\mathrm{c}$ at $z=0$ for the $w$CDM and $f(R)$CDM models respectively.
In order to assess the goodness-of-fit, we compute the residuals normalized by the noise in the data in each mass bin.
The distribution of the normalized residuals for all training cosmological models as a function of mass is also presented in Fig.~\ref{fig:hmf_bspline_fit}.
As we can see, the $97.7$, $84.1$, $50$, $15.9$, $2.3$ percentiles closely match the $1$ and $2\sigma$ levels over the bulk of mass range considered, which is consistent with the expectation of a normal distribution with unit variance.
We find no significant degradation of the quality of the B-spline fit due to the merging of the large mass spline intervals.
Additionally, we detect no discontinuity at the transition between both simulation boxes, which takes place in the mass range $\sim(1-5)\cdot10^{14}\,h^{-1}M_{\odot}$ depending on the particular cosmological model (see Sect.~\ref{subsec:hmf_training_data}).
Only at the low-mass end in the $w$CDM case the residuals are larger than the expected noise of the simulations ($\sim 0.1\%$), nonetheless they remain smaller than $1\%$.
This is because the $w$CDM simulations have $8$ times the number of realizations of the $f(R)$CDM models, hence the statistical noise of the estimated HMFs is systematically smaller.
Therefore, we can conclude that the uncertainties on the B-spline fits are dominated by the noise of the estimated HMF from simulations except some cases at small masses, where uncertainties saturate at $\sim0.5\%$ level.
This shows that the B-spline decomposition accurately describe the HMFs of $w$CDM and $f(R)$CDM cosmologies across the mass range of interest with percent level accuracy.
We find similar results at higher redshifts, as well as for the other halo definitions considered in this work.

\subsection{Emulating the B-spline coefficients}
\label{subsec:bspline_coeff}

The last step to emulate the HMF is to account for the cosmological and redshift dependence of the $c_{j}$ coefficients.
We do so by means of Gaussian process regression.
Gaussian processes have been successfully used in multiple cosmological emulators \citep[e.g.][]{habib_emulator, Coyote_Lawrence_2010, Lawrence_2017, ramachandra_emulator, dark_quest_hmf_emu, Arnold2021, CosmicEmu_IV, emantis_ps_boost}.
We refer the reader to~\citet{gp_rasmussen_williams} for an extensive review of GPs and their application to machine learning.

We fit an independent GP model for each B-spline coefficient, using a Matern-$5/2$ kernel.
We add the variance of the B-spline coefficients to the diagonal of the kernel.
Before fitting the GPs, we standardize the training data and the cosmological parameters, by removing their mean value and rescaling them by their standard deviation.
The standardization of the training data and the GP regression is done with the Python package \textsc{scikit-learn}~\citep{scikit-learn}.
One advantage of GP based emulation, is that the GP is able to provide an uncertainty for each prediction.
Therefore, we can obtain an estimation of the emulation error on each B-spline coefficient.
We can then propagate the uncertainty on each B-spline coefficient prediction to the HMF, by drawing multiple random HMFs and computing their standard deviation.
This way our emulator is able to give an estimation of the emulation error on the predicted HMF, as a function of halo mass, redshift and cosmological parameters.
In Sect.~\ref{subsec:loo_test} we give an assessment of the validity of such estimation.

Our emulation strategy neglects the correlation between B-spline coefficients, which may carry useful information.
To test this, we perform a principal component analysis to decompose the B-spline coefficients into a basis of independent coefficients before fitting the GPs.
However, we find that such procedure does not improve the final accuracy of the emulator.
A more sophisticated multi-output GP regression model such as the one adopted in~\citet{mira_titan_hmf_emu} could be used to include these correlations into the emulation procedure.
This could improve both the accuracy of the emulation and the estimation of the emulator errors.
For the time being, we leave such possibility to future work.
Indeed, in Sect.~\ref{subsec:loo_test} we find that the emulator achieves a few percent accuracy at low masses, and the statistical noise in the simulation data dominates the error budget at high masses.
Furthermore, we find a reasonable agreement between the predicted and the measured emulator errors.

The procedure described above is repeated independently for each of the $19$ redshift outputs, or redshift nodes, for which halo profiles and properties have been saved in the range $0 < z < 2$ (see Sect.~\ref{subsec:nbody_suite}).
Then, in order to obtain predictions at an arbitrary redshift, we perform a linear interpolation of the logarithm of the HMF, $\log_{10} \left[\mathrm{d}n/\mathrm{dln}M_h\right]$, as a function of the scale factor $a = 1/(1+z)$.
A similar strategy is used by \citet{mira_titan_hmf_emu}, although we dispose of $\sim2$ times more redshift nodes from which to interpolate from.
In the rest of this paper, we only perform validations of the emulator predictions up to a maximum redshift of $z_{\rm max}=1.5$, even though the emulator is technically able to give predictions for higher redshifts (see Fig.~\ref{fig:emu_mass_range}).

\section{Validation, tests, and comparisons}
\label{sec:results}

\subsection{Validation with the leave-one-out method}
\label{subsec:loo_test}

\begin{figure*}
	\centering
	\begin{subfigure}{0.49\linewidth}
      \includegraphics[width=0.95\linewidth]{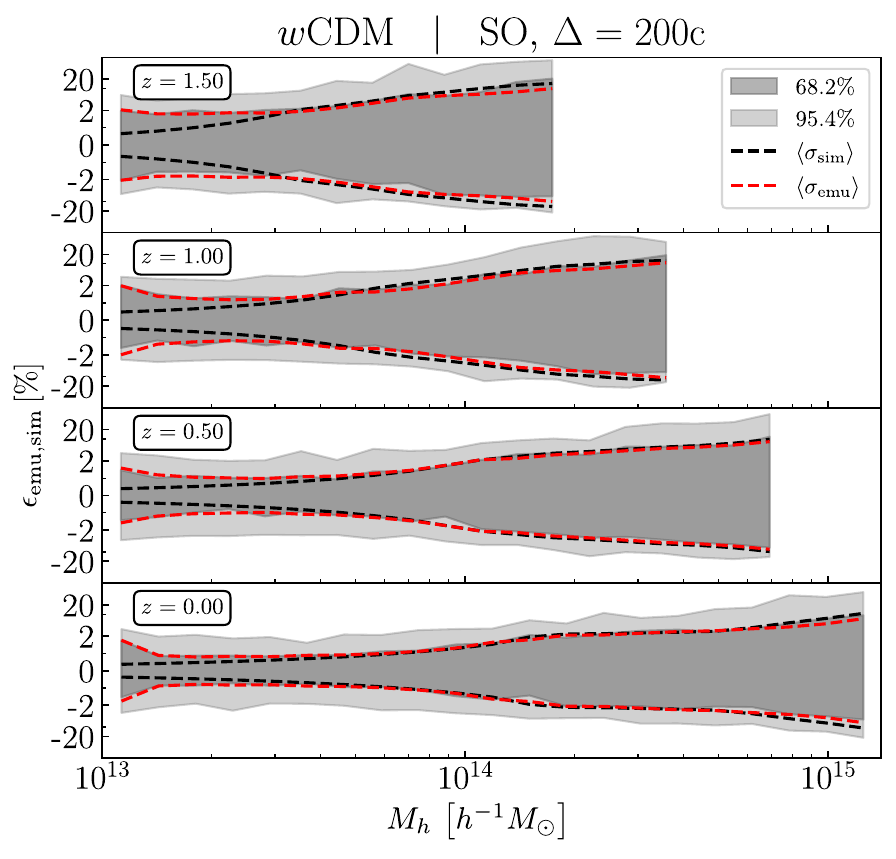}
	\end{subfigure}
	\begin{subfigure}{0.49\linewidth}
		\includegraphics[width=0.95\linewidth]{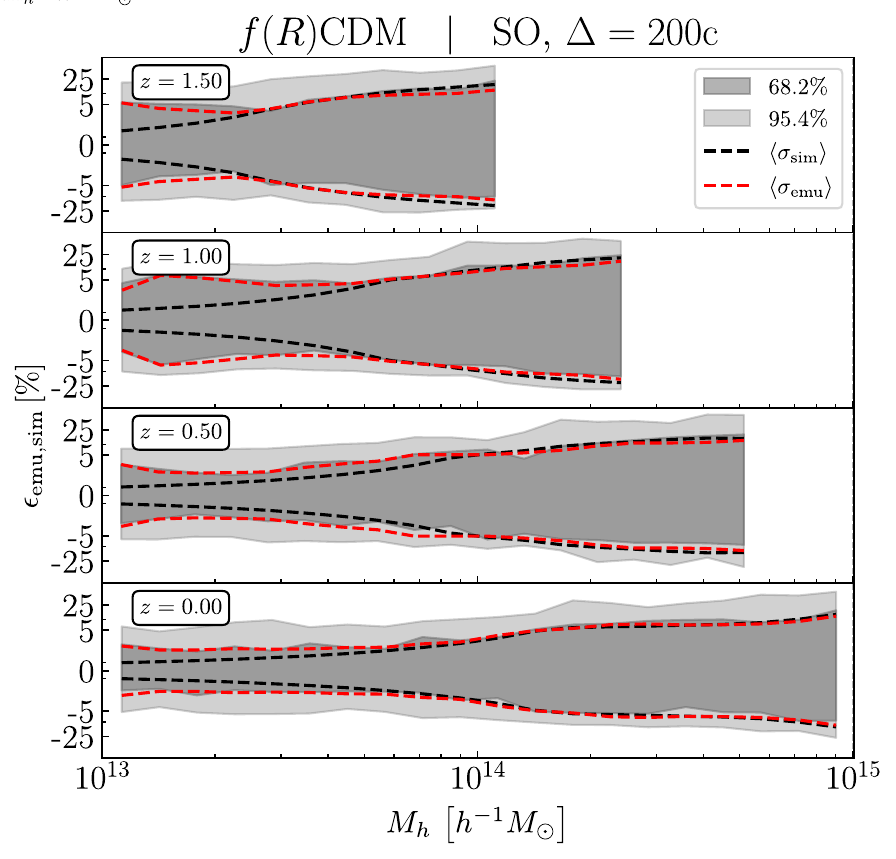}
	\end{subfigure}
	\caption{
    Results of the leave-one-out test for the $w$CDM (\textit{left}) and $f(R)$CDM  (\textit{right}) cosmologies, for multiple redshifts, and in the case of the SO HMF with $\Delta=200\mathrm{c}$.
    One model is excluded from the emulator training set, the corresponding predictions are compared to the simulation measurements of this model.
    The procedure is repeated for each of the $80$ training models.
    The light and dark shaded areas mark the $68.2\%$ and $95.4\%$ confidence intervals of the resulting distribution of relative errors.
    The black dashed line gives the $1\sigma$ error from the simulation measurements, averaged over all cosmological models.
    The red dashed line indicates the $1\sigma$ emulation error estimated by the emulator itself, averaged over all cosmological models.
    In the left plot, the y-axis scaling is linear within the $\pm2\%$ region and logarithmic outside.
    In the right plot, the y-axis scaling is linear within the $\pm5\%$ region and logarithmic outside.
	}
	\label{fig:hmf_loo_deltac200}
\end{figure*}

The leave-one-out test consists in leaving a single model out of the emulator training set, and then comparing the predictions of the emulator to the simulation data of the left-out model.
This is repeated for each model in the emulator training set, thus providing an estimation of the emulator accuracy across the whole parameter space. 

It is worth noting that this is a quite stringent test.
Firstly, uncertainties are tested across the sampled parameter space, including the edges.
Secondly, after removing the tested model, the training set is incomplete and by design the left-out model is far from the other training points.
Hence, if the emulator provides a good description of the left-out model, that is definitely an indication of robustness.
Of course, the final emulator is trained using all the models, which results in smaller emulation errors.
Nevertheless, this leave-one-out test provides a useful conservative estimate of the HMF emulation accuracy.

In Fig.~\ref{fig:hmf_loo_deltac200} we plot the results of the leave-one-out test for the SO HMF with $\Delta=200\mathrm{c}$, and different redshifts, in the case of the $w$CDM and $f(R)$CDM models.
In particular, we show the $68.2\%$ and $95.4\%$ confidence intervals of the distribution of relative errors.
We find that the half width of the $68.2\%$ interval is approximately described by a function of the form $\max\left\{\epsilon_0, \epsilon_0\cdot\left(M_h/M_t\right)^\alpha\right\}$.
In $w$CDM, at $z=0$, we find $\epsilon_0\simeq1.5\%$, $M_t\simeq2\cdot10^{14}\,h^{-1}M_\odot$ and $\alpha\simeq0.9$, up to the maximum halo mass of $M_h\simeq10^{15}\,h^{-1}M_\odot$.
As already shown in Fig.~\ref{fig:emu_mass_range}, the emulator range decreases at higher redshifts.
As expected, the emulation accuracy at a given halo mass decreases with redshift.
In the case of $f(R)$CDM we find at $z=0$: $\epsilon_0\simeq4\%$, $M_t\simeq6\cdot10^{13}\,h^{-1}M_\odot$ and $\alpha\simeq0.4$.
Given the smaller number of $f(R)$CDM realizations per cosmological model, it is not surprising that the emulation accuracy is worse than in the $w$CDM case.

Figure~\ref{fig:hmf_loo_deltac200} also shows the $1\sigma$ errors of the simulation data averaged over all cosmological models.
At large masses, it follows the $68.2\%$ confidence level of the relative errors from the leave-one-out test.
We can conclude that in the high mass regime, the accuracy of the emulator is limited by the statistical noise in the training data.
Increasing the accuracy of the emulator in this regime requires higher volume simulations or a larger number of realizations per cosmological model.

In order to validate the emulator error estimated by the emulator itself, we compare it to the error computed by the leave-one-out test.
We find a good statistical agreement between both errors in the low mass end of the HMF.
However, the emulator tends to overestimate the emulation accuracy in the high mass end.
We multiply the standard deviation predicted by the GPs by fixed factors, such that the distribution of leave-one-out errors normalized by the emulator estimated errors is close to a unit-variance normal distribution.
This kind of technique to correct the error predicted by the emulator itself has already been used by \citet{mira_titan_hmf_emu}.
We give in Fig.~\ref{fig:hmf_loo_deltac200} the corrected $1\sigma$ emulation error, as estimated by the emulator itself, and averaged over all cosmological models.
It can be seen that this error is in good agreement with the $68.2\%$ confidence level of the leave-one-out errors for all masses and redshifts, which serves as a validation of the uncertainty predicted by the emulator.
It means that the emulator is able to estimate its own accuracy, enabling users to account for (often overlooked) theoretical error bars during the cosmological inference process.
In particular, the errors in the low mass end of the HMF are dominated by the interpolation between cosmological models and not the noise in the simulation data.
Improving the emulator accuracy at low masses would require a higher number of training models or a more efficient emulation procedure.
To conclude, the \textsc{e-mantis} emulator predicts the HMF at the few percent level over a wide range of masses and redshifts in $f(R)$CDM and $w$CDM cosmologies.

\subsection{Test with other \texorpdfstring{$N$}{N}-body simulations}
\label{subsec:comparison_to_nbody}

\subsubsection{Test with e-MANTIS reference simulations}
\label{subsubsec:comparison_to_emantis_sims}

\begin{figure*}
	\centering
	\begin{subfigure}{0.49\linewidth}
		\includegraphics[width=0.95\linewidth]{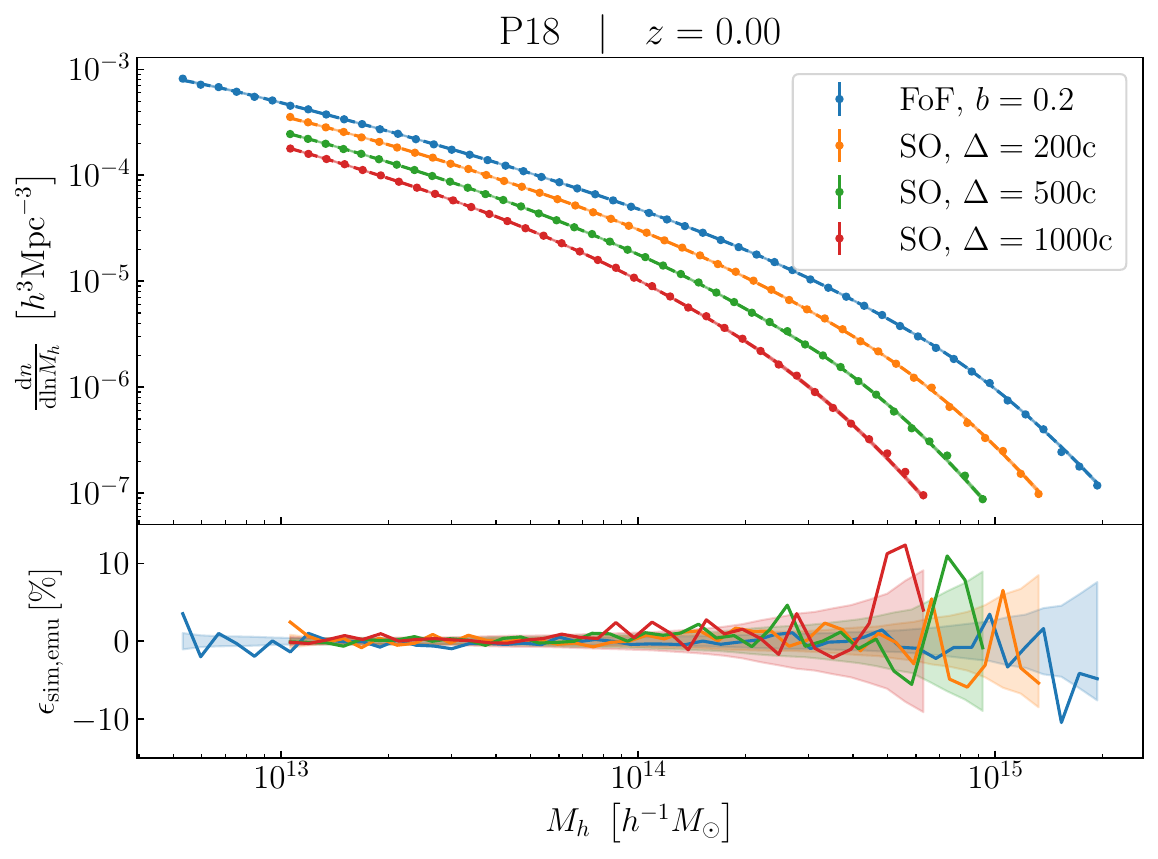}
	\end{subfigure}
	\begin{subfigure}{0.49\linewidth}
		\includegraphics[width=0.95\linewidth]{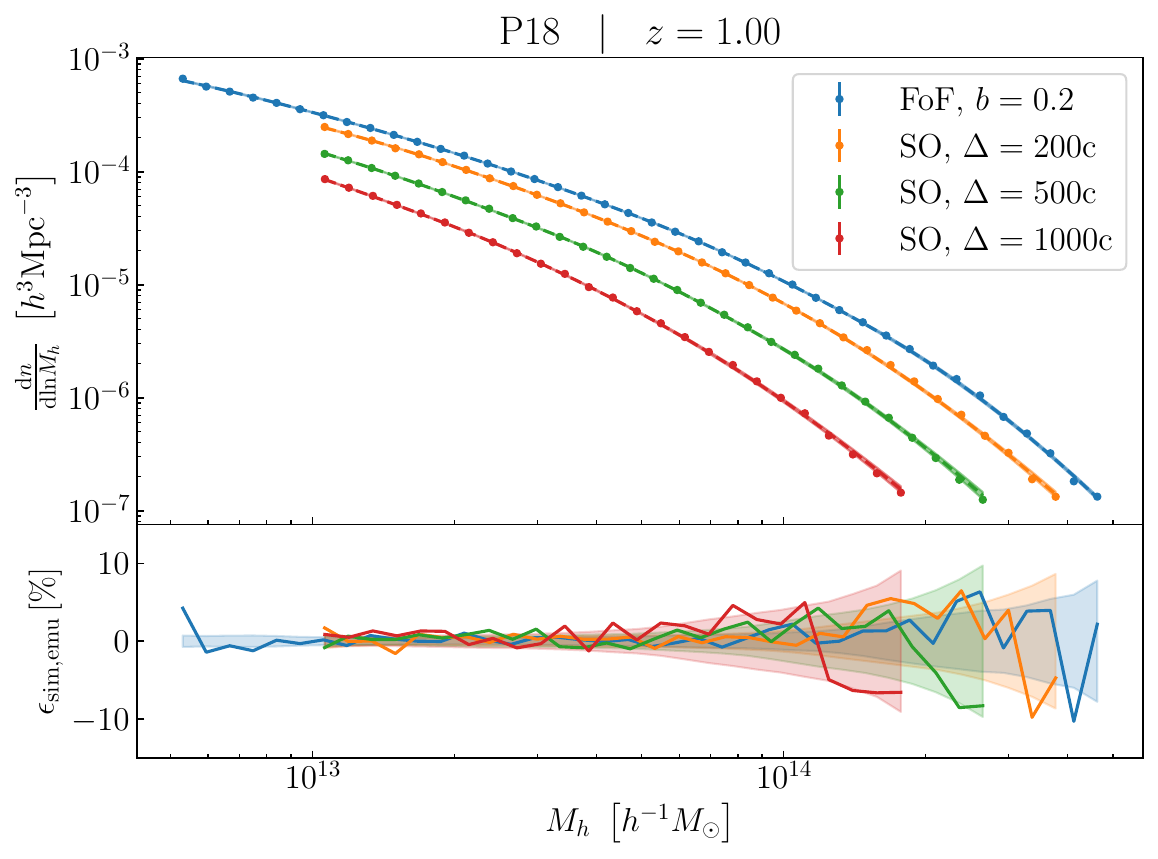}
	\end{subfigure}
	\caption{
    Comparison of the emulator HMF predictions to the L$328\_$M$10\_$P$18$ simulation at $z=0$ (\textit{left}) and $z=1$ (\textit{right}).
    Top panels: HMFs measured from the simulation (solid circles) and emulator predictions (dashed lines) without resolution correction, such as to consistently match the resolution of the L$328\_$M$10\_$P$18$ simulation.
    The different colours correspond to the different halo mass definitions.
    The error bars correspond to the HMF simulation measurement errors, while the shaded area around the lines correspond to the uncertainties of HMF emulator predictions (not visible due to the small level of uncertainty).
    Bottom panels: relative difference between the simulation measurements and the emulator predictions.
    The shaded area correspond to the $1\sigma$ confidence interval combining both the simulation noise and the emulator accuracy.
    There is a good agreement with differences of the order of a few percent level, and within the expected error bars.
	}
	\label{fig:emu_vs_sim_emantis_p18}
\end{figure*}

\begin{figure*}
	\centering
	\begin{subfigure}{0.49\linewidth}
		\includegraphics[width=0.95\linewidth]{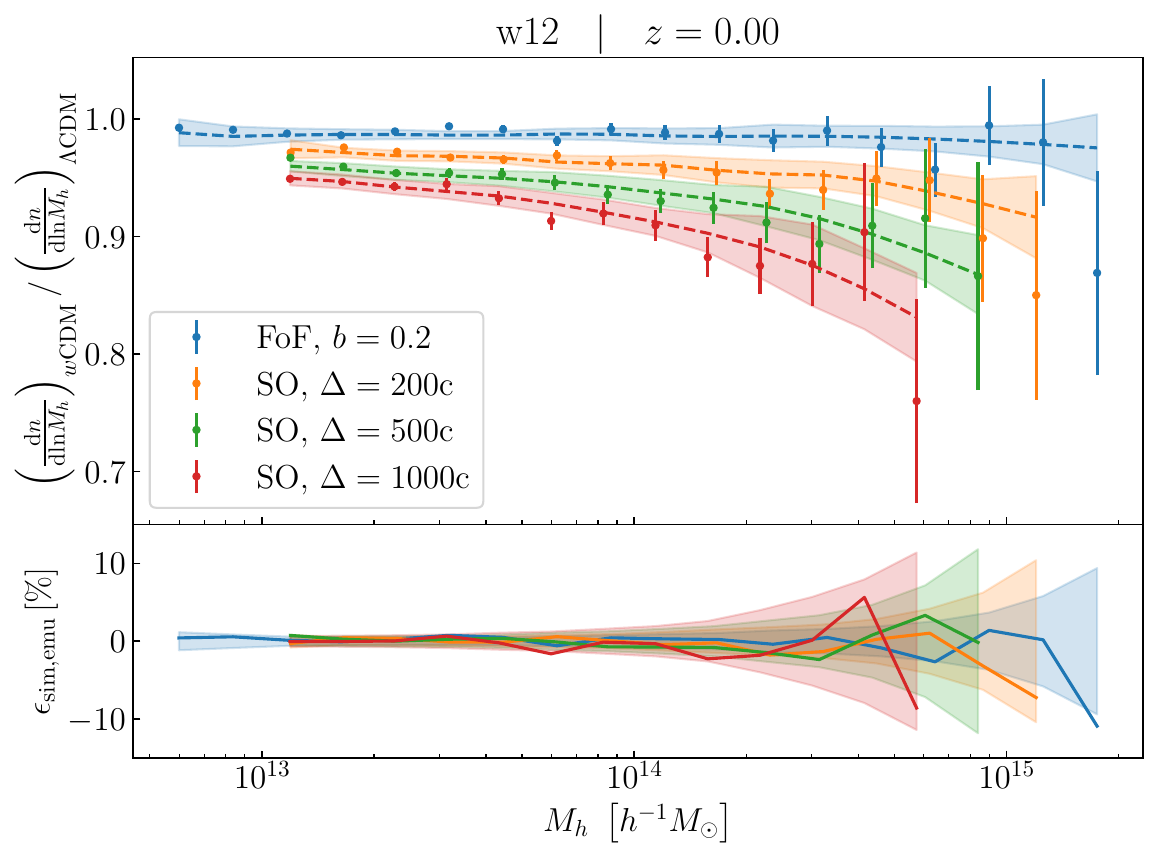}
	\end{subfigure}
	\begin{subfigure}{0.49\linewidth}
		\includegraphics[width=0.95\linewidth]{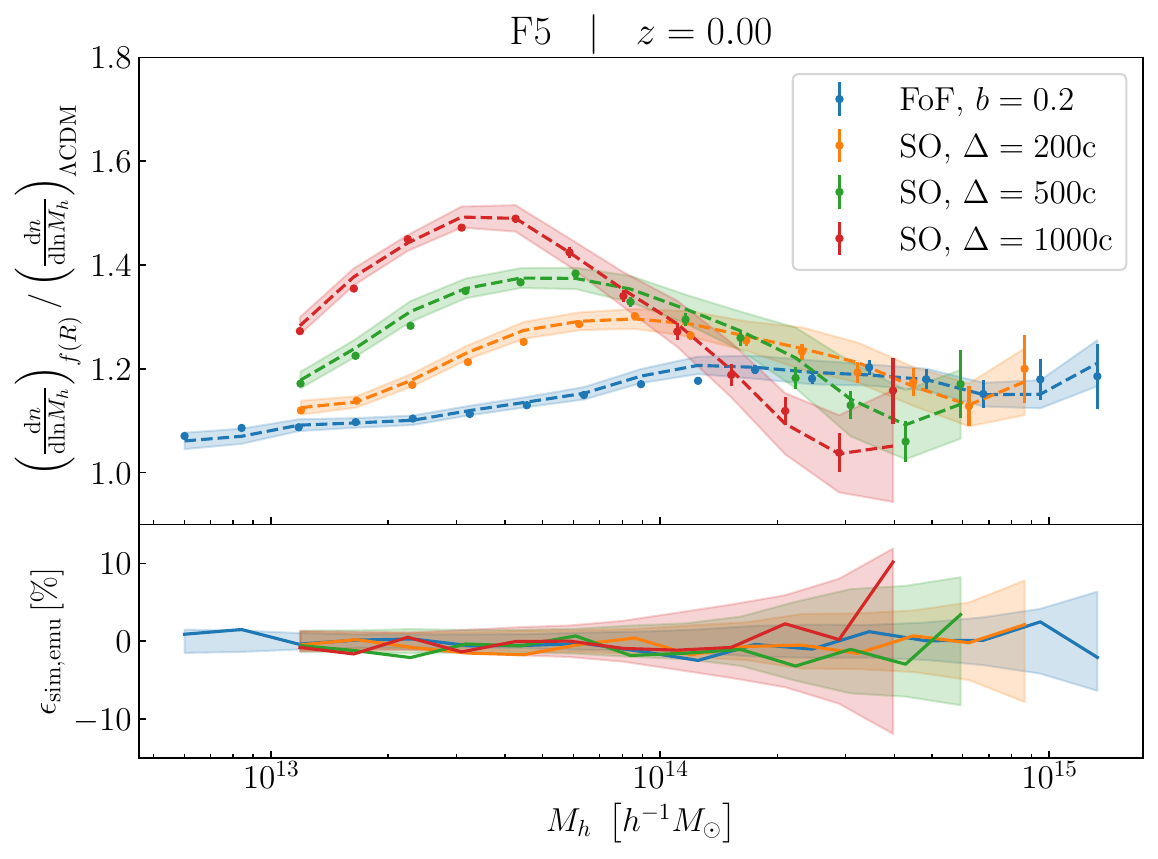}
	\end{subfigure}
	\caption{
    Comparison of the emulator HMF predictions at $z=0$ in $w$CDM (\textit{left}) and $f(R)$CDM (\textit{right}) to the HMF of L$328\_$M$10\_$w$12$ (\textit{left}) and L$328\_$M$10\_$F$5$ (\textit{right}).
    The HMFs are normalized by the L$328\_$M$10\_$P$18$ simulation data.
    Top panels: Ratio of the $w$CDM (\textit{left}) HMF or $f(R)$CDM (\textit{right}) HMF to the $\Lambda$CDM one measured from the simulation (solid circles) and from emulator predictions (dashed lines).
    The error bars correspond to the HMF simulation measurement errors (not always visible for small errors), while the shaded area around the  lines correspond to the uncertainties of HMF emulator predictions.
    Bottom panels: relative difference between the HMF simulation measurements and the emulator predictions.
    The shaded area correspond to the $1\sigma$ confidence interval combining both the simulation noise and the emulator accuracy.
    There is a good agreement with differences of the order of a few percent level, and within the expected error bars.
    }
	\label{fig:emu_vs_sim_emantis_wcdm_fofR}
\end{figure*}

To test the predictive power of the HMF emulator, we compare its predictions to the set of simulations for the reference cosmological models P$18$, w$12$ and F$5$ presented in Sect.~\ref{subsec:nbody_suite}.
The P$18$ and w$12$ cosmological models are chosen to be within $2\sigma$ of the cosmological parameters inferred from the 2018 analysis of \textit{Planck} CMB data \citep{Planck2018}.
The F$5$ model represents a significant deviation from $\Lambda$CDM, while being compatible with recent cluster abundance constraints~\citep{Artis_2024}. 
These models allow us to gauge the efficiency of the emulator in cosmologies which at the time of writing are considered realistic.
These simulations use the $\mathrm{L328}\_\mathrm{M}10$ box, and have $384$ independent realizations for P$18$ and $64$ realizations for w$12$ and F$5$.
As these data were not used to train the emulator, they provide an ideal reference to which we can compare the predictions of the emulator.

In Fig.~\ref{fig:emu_vs_sim_emantis_p18} we plot the HMF predicted by the emulator against that measured from the L$328\_$M$10\_$P$18$ simulation, at $z=0$ and $z=1$, and for the different halo mass definition.
For this comparison, we use \textsc{e-mantis} without the low mass end resolution correction (see Sect.~\ref{subsec:mass_res_corr}) in order to match the M$10$ mass resolution of the L$328\_$M$10\_$P$18$ simulation.
In the top panels, we can see that there is a good agreement between the emulated HMFs and the simulation data, for all the halo definitions supported by the emulator.
The bottom panels show the relative difference between the measured HMF and the predicted one, with the shaded area indicating the overall uncertainty estimated by adding in quadrature the error of the estimated HMF from the reference simulations (assumed to be Poisson noise) and the emulator uncertainty.
We find such uncertainties to be dominated by the Poisson error, \corr{which} is related \corr{to} the number of halos in each mass bin.
In particular, we can see that the relative difference between the measured HMF and the predicted one remains at the $1\%$ level across the bulk of the mass interval, for the different halo mass definition.
It increases up to the $\sim 5\%$ level only at the high-mass end, and with a scatter dominated by the Poisson noise, that is due to the reduced number of halos in the highest mass bins.

In Fig.~\ref{fig:emu_vs_sim_emantis_wcdm_fofR}. we compare the HMF predicted by the emulator against the HMF measured from the L$328\_$M$10\_$w$12$ and L$328\_$M$10\_$F$5$ simulations, at $z=0$.
More specifically, in the top panels we show the ratio of the HMF with respect to that of the L$328\_$M$10\_$P$18$ simulation, while in the bottom panels we show the relative difference between the simulations measurements and the emulator predictions.
Once again, there is a good agreement between the predictions of the HMF emulator and the HMFs from simulations.
In the w$12$ case, the differences are within the $1\%$ level across the bulk of the mass interval, with larger errors occurring only in the highest mass bins due to Poisson noise.
We obtain similar results for the F$5$ case, with differences at the $2\%$ level over most of the mass interval.
In both cases, the differences are within the expected combined uncertainties from the emulator and the shot-noise from the simulation data.
The fact that the uncertainties estimated from the F$5$ analysis are slightly larger than the w$12$ case is consistent with the fact that in the former case, the training set contains a factor 8 fewer simulations per cosmological model.
Indeed, as explained in Sect.~\ref{subsec:loo_test}, the emulator error in $f(R)$CDM is a bit larger than in $w$CDM.

\begin{figure*}
	\centering
    \begin{subfigure}{0.49\linewidth}
      \includegraphics[width=0.95\linewidth]{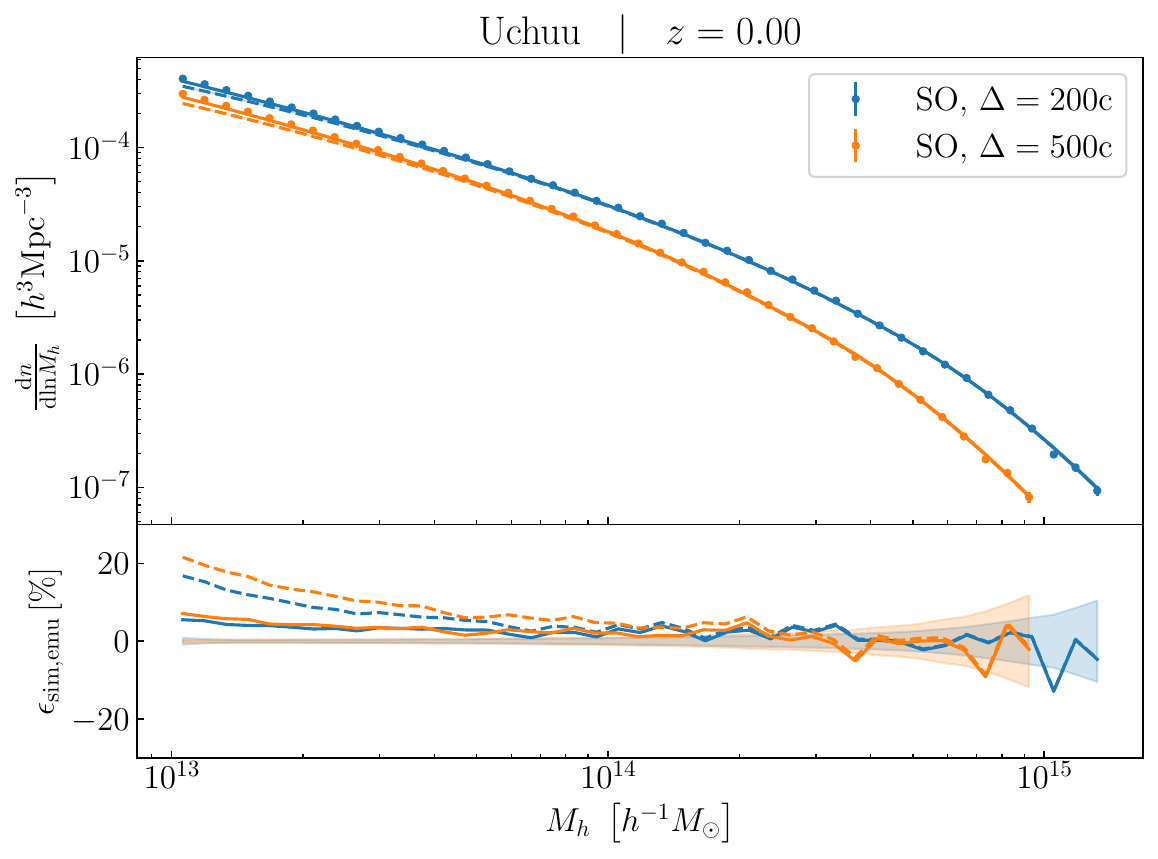}
    \end{subfigure}
    \begin{subfigure}{0.49\linewidth}
      \includegraphics[width=0.95\linewidth]{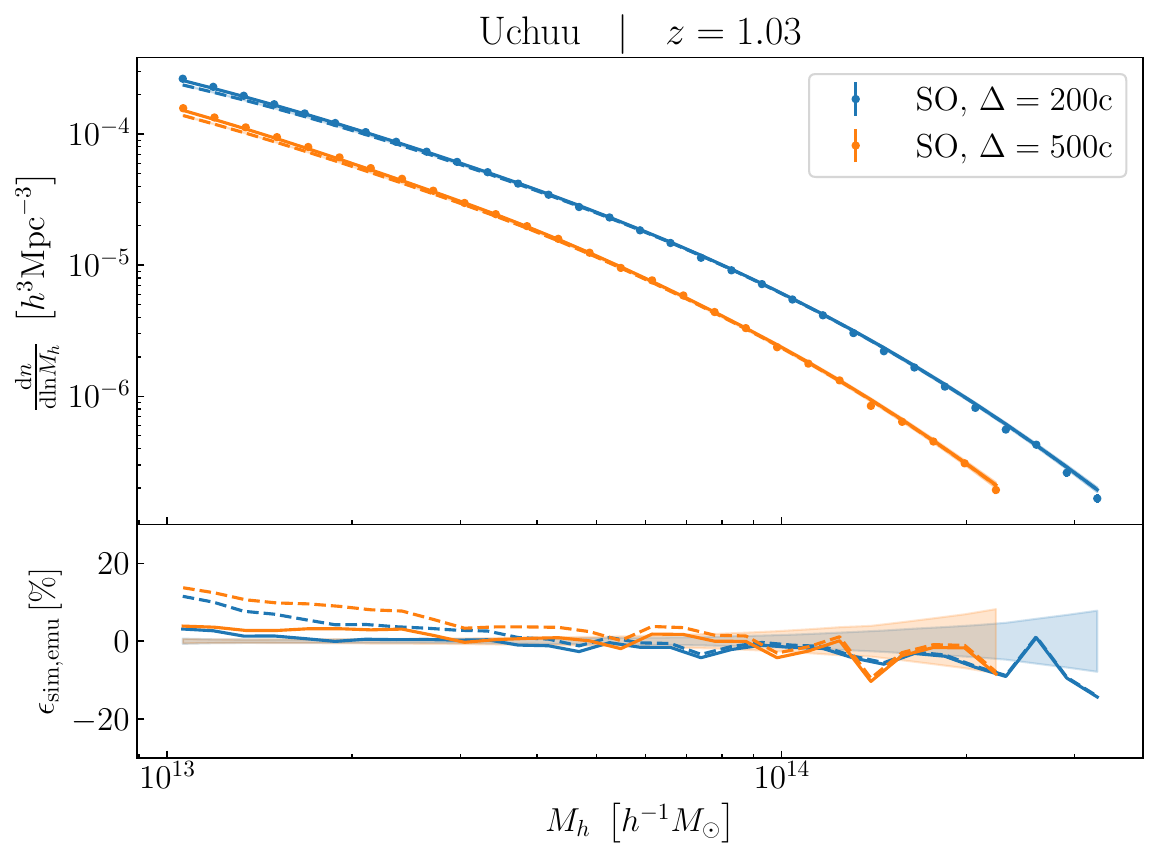}
    \end{subfigure}
	\caption{
    Comparison of the emulator HMF predictions to that estimated from the Uchuu simulation, \corr{at $z=0$ (\textit{left}) and $z=1.03$ (\textit{right})}.
    Top panels: HMF from the simulation (solid circles) and emulator predictions with (solid lines) and without (dashed lines) resolution correction.
    The noise in the simulation measurements and the emulator predictions are given as error bars and a shaded area respectively around the corresponding lines (not always visible for small errors).
    The Uchuu haloes have been detected with the \textsc{rockstar} halo finder~\citep{Behroozi2013}.
    Bottom panels: relative difference between the simulation measurements and the emulator predictions with (solid lines) and without (dashed lines) resolution correction.
    The shaded areas mark the $1\sigma$ confidence interval, combining both the simulation data noise and the emulator accuracy.
	}
	\label{fig:emu_vs_sim_uchuu}
\end{figure*}

Overall, we find the emulator to be in good agreement with the simulation set for the reference P$18$, F$5$ and w$12$ cosmological models.
In particular, there is a few per cent level agreement over a wide range of halo masses and for multiple halo definitions.
At large masses, the differences grow due to the statistical noise both in the reference simulation data and in the emulator training set.
In any case, the differences between the emulator predictions and the simulation data remain within the expected error bars.

\subsubsection{Test with publicly available \texorpdfstring{$N$}{N}-body simulations}
\label{subsubsec:comparison_to_other_sims}

We now present a comparison of the HMF predicted by the \textsc{e-mantis} emulator against numerical estimates from publicly available $N$-body simulations.
First, we use the halo catalogues from the Uchuu simulation\footnote{\href{https://www.skiesanduniverses.org/Simulations/Uchuu/}{https://www.skiesanduniverses.org/}} \citep{Ishiyama2021}.
These allow us to compare the HMF prediction of the emulator to a simulation which has larger volume, ($2\ h^{-1}$Gpc)$^3$, and a better mass resolution $m_{\rm part} = 3.28\cdot10^{8} h^{-1}{\rm M}_{\odot}$, than those of the \textsc{e-mantis} training set.
Moreover, the Uchuu simulation has been run using a different numerical $N$-body code from the one used in this work, \textsc{greem}~\citep{Ishiyama2009, Ishiyama2012}; similarly, the halo catalogues have been generated using a different halo finder, \textsc{rockstar} \citep{Behroozi2013}.
All these differences allow us to test resolution effects as well as the consistency of the results with respect to the different numerical codes used to perform and analyse the \textsc{e-mantis} simulations.
The Uchuu simulation uses a flat $\Lambda$CDM cosmological model compatible with \citet{2016A&A...594A..13P}: $\Omega_{\rm m}=0.3089$, $h=0.6774$, $\sigma_{8}^{\rm GR}=0.8159$, $\Omega_{\rm b}=0.0486$, $n_{\rm s}=0.9667$.

We estimate the Uchuu HMFs at two overdensity contrasts, $\Delta = 200\mathrm{c}$ and $\Delta = 500\mathrm{c}$ respectively.
We do not include the case of $\Delta=1000\mathrm{c}$ in this comparison, since the necessary Uchuu data is not available.
In Fig.~\ref{fig:emu_vs_sim_uchuu} we compare the HMF predicted by the \textsc{e-mantis} emulator to the Uchuu HMFs, with and without the low mass end resolution correction presented in Sect.~\ref{subsec:mass_res_corr}, \corr{both at $z=0$ and $z=1.03$}.
We can see that including the correction significantly reduces the relative difference between the emulator predictions and the simulation measurements from $\sim 20\%$ to $\sim 5\%$ at the low mass end ($M_{h} \simeq 10^{13}\ h^{-1}{\rm M}_\odot$) \corr{and at $z=0$}, showing that the emulator predictions are in agreement with the Uchuu HMFs at the few percent level, despite the many differences between the simulation settings.
We attribute the remaining differences to the different halo finders \citep{Knebe2011} and $N$-body codes.
\corr{We note that the discrepancies in the low mass end are smaller at $z=1.03$.}
In the high mass end, the differences are within the expected error bars from the emulator uncertainty combined with the simulation shot-noise.
\corr{We find similar results for all the other redshifts in the range $0 < z < 1.5$.}

\begin{figure*}
	\centering
    \begin{subfigure}{0.49\linewidth}
      \includegraphics[width=0.95\linewidth]{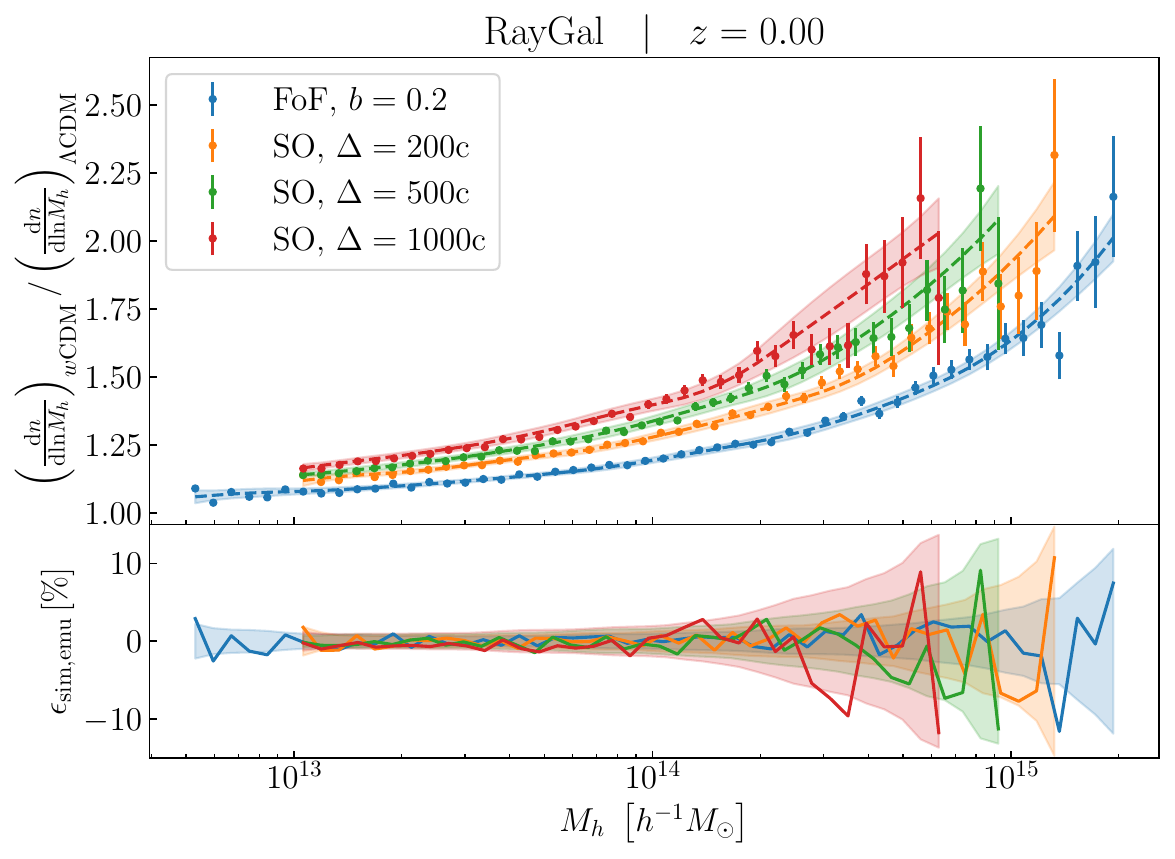}
    \end{subfigure}
    \begin{subfigure}{0.49\linewidth}
      \includegraphics[width=0.95\linewidth]{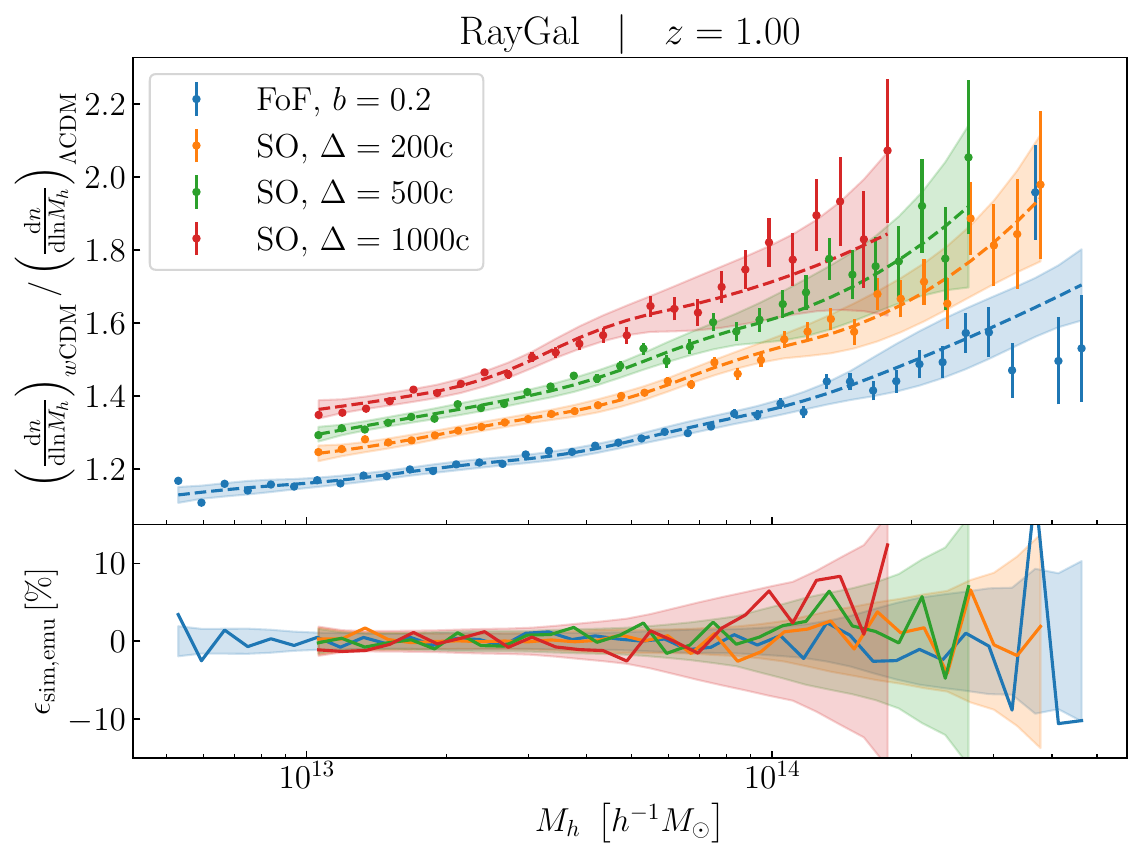}
    \end{subfigure}
	\caption{
      Comparison of the \textsc{e-mantis} predictions for the ratio of the HMFs from the two RayGal simulations for different halo definitions at $z=0$ \corr{(\textit{left}) and $z=1$ (\textit{right})}.
      Both RayGal simulations differ not only in the value of $w$, but also in terms of $\Omega_{\rm m}$ and $\sigma_8^{\rm GR}$.
      Top panel: ratio of the HMFs from the two RayGal simulations (solid circles) and predictions from the emulator (dashed lines).
      The noise in the simulation measurements and the emulator predictions are given as error bars and a shaded area respectively around the corresponding lines (not always visible for small errors).
      Bottom panel: relative difference between the simulation measurements and the emulator predictions.
      The shaded area marks the $1\sigma$ confidence interval, combining both the simulation data noise and the emulator accuracy.
	}
	\label{fig:emu_vs_sim_raygal}
\end{figure*}

To benchmark the predictive power of the emulator for cosmologies alternative to $\Lambda$CDM, we consider two additional simulation suites, the $w$CDM \textsc{RayGalGroupSims} runs \citep{Rasera2022}, hereafter RayGal, and the $f(R)$ gravity \textsc{elephant} runs \citep{Cautun2018, Gupta_2022}.
Both these suites adopt numerical settings similar  to those of the \textsc{e-mantis} simulation training set.
As an example, both suites were run using the code \textsc{ramses}.
Furthermore, the \textsc{elephant} suite also uses the \textsc{ecosmog} implementation \citep{Li2012, Bose2017} of the \cite{2007PhRvD..76f4004H} $f(R)$ modified gravity model.
\corr{
  Here we focus on the HMF boost, i.e. the ratio of the HMF in an alternative cosmology with respect to $\Lambda$CDM.
  Different systematic errors, such as resolution effects or differences in the halofinders, largely cancel out using this boost.
  We discuss some of these systematic effects in more details in Sect.~\ref{subsubsec:other_preds_lcdm}.
  The aim of the present test is to focus on the cosmological dependence with respect to the alternative parameters.
}

The RayGal runs consist of two large volume simulations of $(2625 h^{-1}{\rm Mpc})^3$ of a flat $\Lambda$CDM model and a flat $w$CDM scenario with constant equation of state $w=-1.2$.
The cosmological parameters of these simulations were chosen to be consistent with the results of the WMAP-7 year \citep{WMAP7} and \textit{Planck} \citep{Planck2018} data of the CMB anisotropy power spectra.
In particular, the $\Lambda$CDM run assumes $\Omega_{\rm m}=0.25733$ and $\sigma_8^{\rm GR} = 0.80101$, while the $w$CDM run assumes $\Omega_{\rm m} = 0.27508$ and $\sigma_8^{\rm GR} = 0.85205$.
The remaining cosmological parameters are fixed to $\Omega_{\rm b} = 0.04356$, $h = 0.72$, $n_{\rm s} = 0963$, and $\Omega_{\rm r} \simeq 0.00008$ respectively.
In both simulations, the dark matter density field is traced by $4096^3$ $N$-body particles, leading to the two simulations having slightly different mass resolution, with $m_{\rm part} = 1.88 \cdot10^{10}\,h^{-1}{\rm M}_\odot$ in the case of the $\Lambda$CDM run and $m_{\rm part} = 2.01 \cdot10^{10}\,h^{-1}{\rm M}_\odot$ for the $w$CDM case.

It is worth noticing that for each cosmological model in the $w$CDM emulator training dataset, the sample of small volume simulations, $\mathrm{L}328\_\mathrm{M}10\_\mathrm{wcdm}$, represent a total effective volume equivalent to an eighth of the full RayGal volume for each training cosmological model, with similar mass resolution (not considering the resolution correction presented in Sect.~\ref{subsec:mass_res_corr}).
We note that the RayGal simulations used a Triangular Shaped Cloud (TSC) interpolation in the dynamical solver, while the \textsc{e-mantis} training set uses a Cloud in Cell (CIC) interpolation.
This difference in the interpolation scheme results in a slightly different effective resolution at equal $N$-body particle mass.
The sample of larger volume simulations, $\mathrm{L}656\_\mathrm{M}11\_\mathrm{wcdm}$, covers a volume equivalent to the simulation box of the RayGal runs for each cosmological model, with lower mass resolution.

In Fig.~\ref{fig:emu_vs_sim_raygal} we plot the ratio of the HMF estimated from the $w$CDM RayGal run to the HMF estimated from the $\Lambda$CDM simulation and compare these estimates to the same ratio predicted by \textsc{e-mantis}, \corr{both at $z=0$ and $z=1$}.
We find that the predictions are accurate to the level of a few percent at the low mass end, and reach a maximum difference with respect to the simulation data of $\sim 10\%$ in the largest mass bins.
At all masses, the relative difference is generally consistent with the uncertainty of the emulator and the Poisson noise in the simulation data.
\corr{Similar results are found at other redshifts in the range $0 < z < 1.5$.}
We may notice that the trend shown in Fig.~\ref{fig:emu_vs_sim_raygal} differs from that in Fig.~\ref{fig:emu_vs_sim_emantis_wcdm_fofR}.
This is essentially due to the differences in the values of $\sigma_8^{\rm GR}$ and $\Omega_{\rm m}$ between the two RayGal runs, whereas in Fig.~\ref{fig:emu_vs_sim_emantis_wcdm_fofR} the difference is solely due to the dark energy equation of state parameter.

The \textsc{elephant} runs evolve $1024^{3}$ $N$-body particles in a volume of $(1024\ h^{-1}{\rm Mpc})^3$, with a particle mass of $m_{\rm part}=7.798\cdot10^{10}\,h^{-1}M_{\odot}$.
In addition to a $\Lambda$CDM model, there are simulations for three different $f(R)$ gravity models: $f_{R_{0}} = -10^{-4}$ (F$4$), $f_{R_{0}}=-10^{-5}$ (F$5$) and $f_{R_{0}}= -10^{-6}$ (F$6$).
The other cosmological parameters are fixed to the best-fit values from the WMAP-9 year analysis \citep{WMAP9}: $\Omega_{\rm m} = 0.281$, $\sigma_8^{\rm GR} = 0.820$ $\Omega_\Lambda = 0.719$, $h = 0.697$, $\Omega_{\rm b}=0.046$ and $n_{\rm s} = 0.971$.
There are $2$ independent realizations in F$4$ and $5$ independent realizations in $\Lambda$CDM, F$5$, and F$6$ \citep[see][]{Gupta_2022}.
These simulations are run using the \textsc{ecosmog} implementation of the \cite{2007PhRvD..76f4004H} model.
The dark matter halo catalogues and HMF measurements have been obtained by~\citet{Gupta_2022} using the \textsc{rockstar} halo finder.

\begin{figure*}
	\centering
    \begin{subfigure}{0.49\linewidth}
      \includegraphics[width=0.95\linewidth]{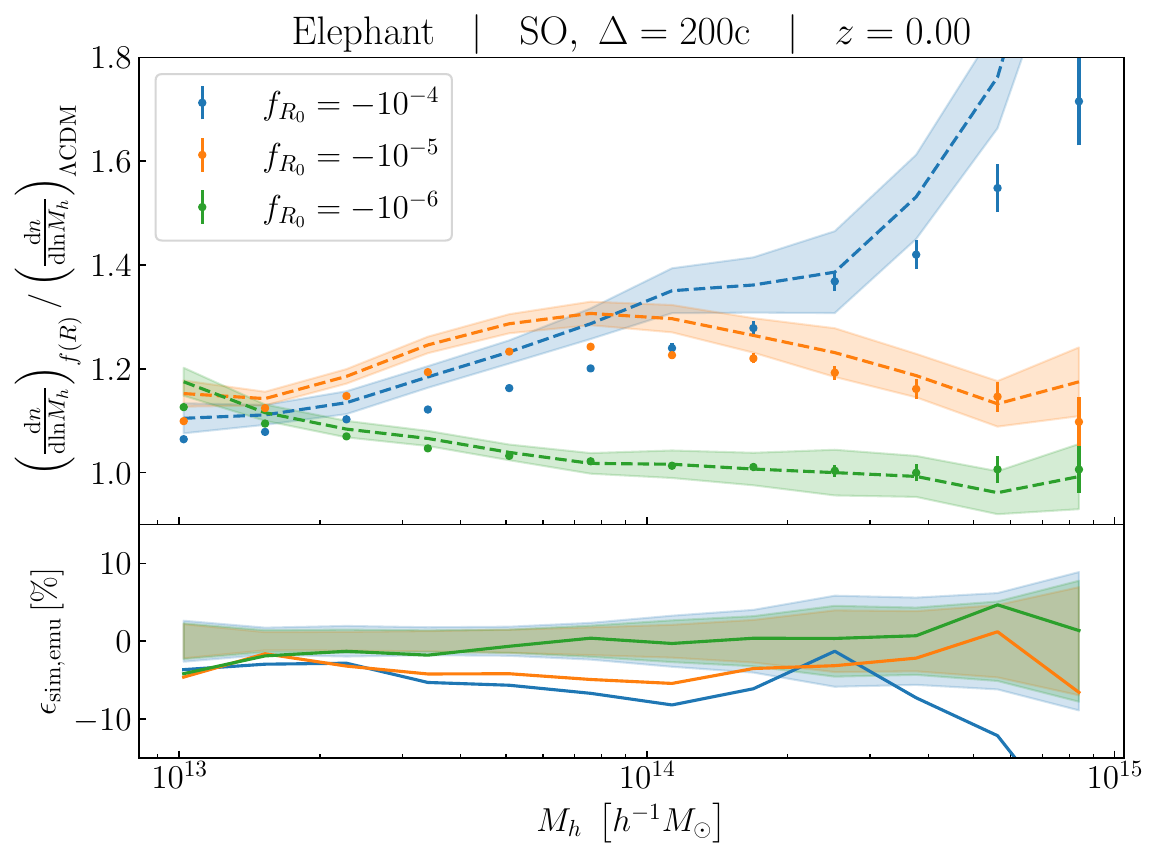}
    \end{subfigure}
    \begin{subfigure}{0.49\linewidth}
      \includegraphics[width=0.95\linewidth]{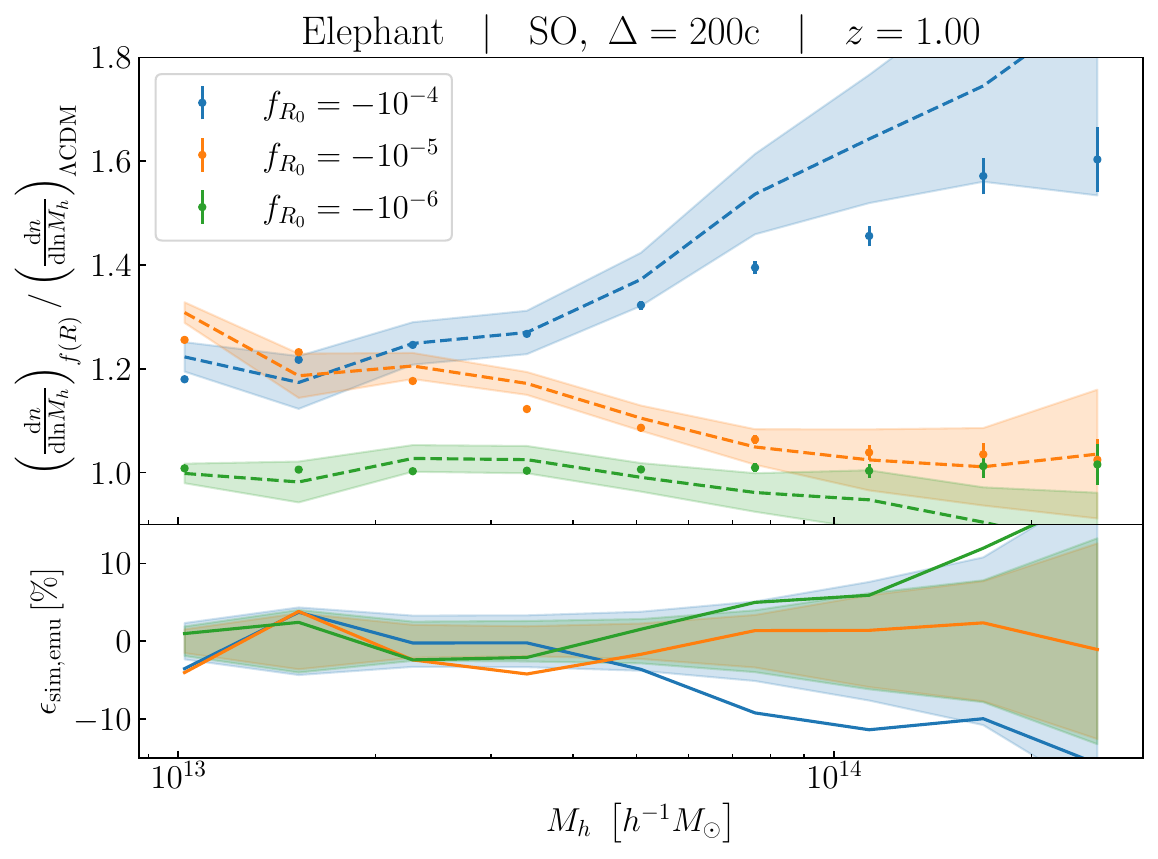}
    \end{subfigure}
	\caption{
      Comparison between the \textsc{e-mantis} predictions and the measurements from the \textsc{elephant} simulations~\citep{Cautun2018, Gupta_2022} for SO haloes with $\Delta=200\mathrm{c}$ at $z=0$ \corr{(\textit{left}) and $z=1$ (\textit{right})}.
      Top panels: measured HMF boost from the simulations (solid circles) and predictions from the emulator (dashed lines).
      The noise in the simulation measurements and the emulator predictions are given as error bars and a shaded area respectively around the corresponding lines (not always visible for small errors).
      Bottom panels: relative difference between the simulation measurements and the emulator predictions.
      The shaded areas mark the $1\sigma$ confidence interval, combining both the simulation data noise and the emulator accuracy.
	}
	\label{fig:emu_vs_sim_elephant}
\end{figure*}

In Fig.~\ref{fig:emu_vs_sim_elephant} we show the HMF ratio in each $f(R)$ model with respect to $\Lambda$CDM, as measured from the \textsc{elephant} simulations and as predicted by \textsc{e-mantis}, \corr{both at $z=0$ and $z=1$}.
In the case of F$6$ at $z=0$, the agreement is within the expected errors from the emulator and the simulation Poisson noise.
\corr{
  At $z=1$, this is also true for masses $M_{h} \lesssim 10^{14}\,h^{-1}\,M_{\odot}$.
  For F$5$, the differences are within the expected error bars at $z=1$, while at $z=0$ they are slightly larger, but remain at the $\sim5\%$ level for the whole mass range.
}
\corr{
  For F$4$, the differences are at the $\sim5\%$ level for masses $M_{h} \lesssim 3\cdot10^{14}\,h^{-1}M_{\odot}$ and $M_{h} \lesssim 5\cdot10^{13}\,h^{-1}M_{\odot}$, for $z=0$ and $z=1$, respectively.
}
These differences might once again be due to the different halo finders, however a more dedicated study is required to reach a conclusion.
\corr{Overall, we find that the predictions of \textsc{e-mantis} for the HMF boost in $f(R)$ gravity are in good agreement with the data from the \textsc{elephant} simulations at the $\sim5\%$ level, for most masses, redshifts, and models, while larger discrepancies are present in some cases, such as for high redshifts together with large masses or large values of $\log_{10}{|f_{R_{0}}|}$.}

\subsection{Comparison to existing HMF predictions}
\label{subsec:comparison_to_other_preds}

In the previous section, we have validated the \textsc{e-mantis} HMF emulator from group-masses to cluster-masses and redshifts $z<1.5$, using both internal and external simulations in $\Lambda$CDM, $w$CDM and $f(R)$CDM cosmologies.
As the goal of \textsc{e-mantis} is to provide fast ($\sim\mathrm{10~ms}$) predictions of the HMF (as well as other observables of the non-linear clustering of matter) for various applications requiring speed and accuracy, such as MCMC cosmological inference, it is of interest to compare its predictions to that of other emulators as well as those from analytically calibrated models of the HMF.
For this purpose, we first compare predictions for the $\Lambda$CDM cosmology, and then discuss the case of $w$CDM and $f(R)$CDM models. 
\corr{
  More specifically, we start by a comparison of the full HMF for a fixed cosmology in $\Lambda$CDM.
  They aim is to study the systematic differences that exist between different predictions for a fixed cosmological model.
  Then, we turn to the study of the HMF boost in $w$CDM and $f(R)$CDM models.
  This time, the objective is to focus on the cosmological dependence.
  As already stated earlier, several systematic errors cancel out in the boost, which allows a more precise study of the cosmological dependence.
}

\begin{figure*}
	\centering
	\begin{subfigure}{0.49\linewidth}
		\includegraphics[width=0.95\linewidth]{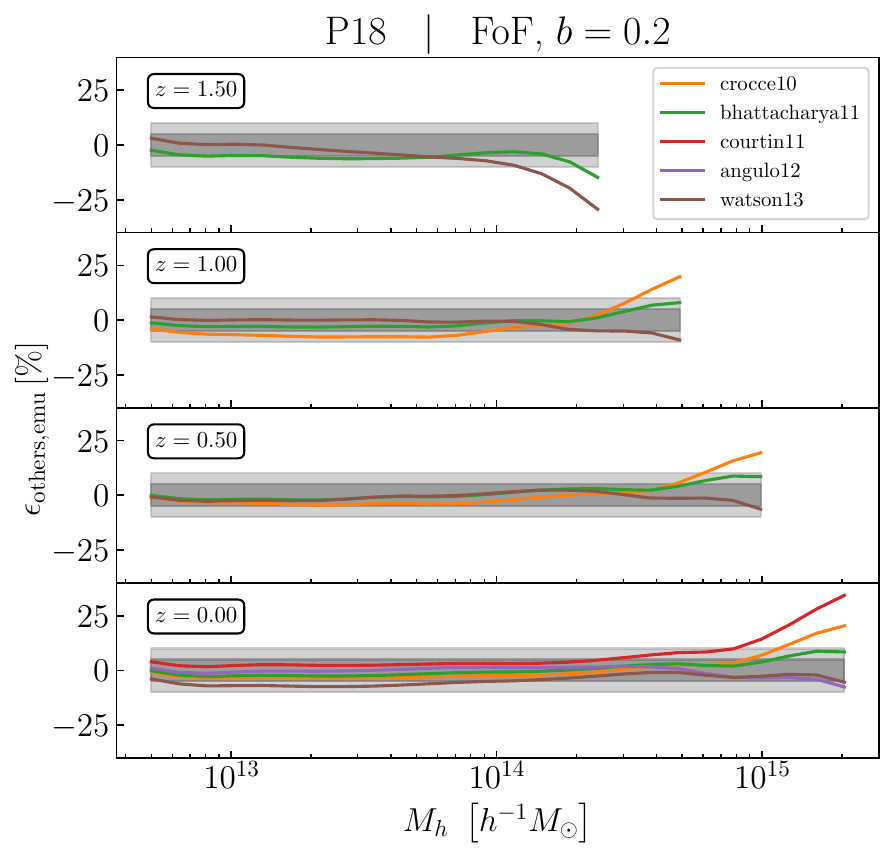}
	\end{subfigure}
	\begin{subfigure}{0.49\linewidth}
		\includegraphics[width=0.95\linewidth]{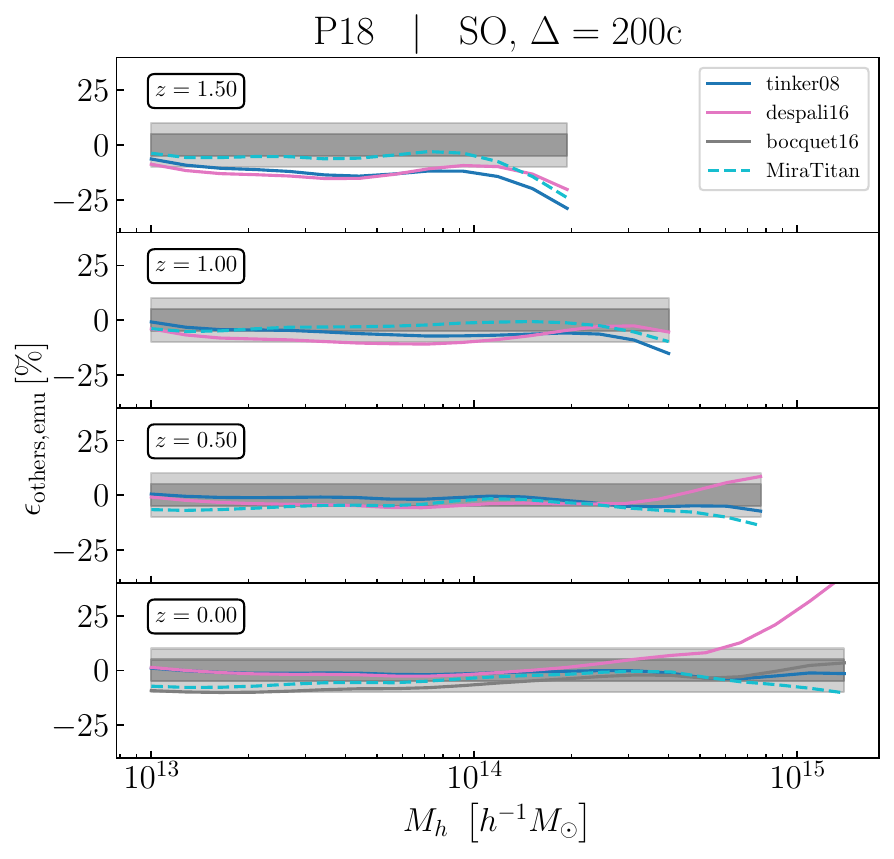}
	\end{subfigure}
	\caption{
    Relative difference between different HMF predictions from other works found in the literature and \textsc{e-mantis}, for the reference P18 cosmology and at different redshifts.
    Solid lines correspond to predictions from fits to simulations, and dashed lines to emulators.
    \textit{Left}: FoF haloes with $b=0.2$.
    \textit{Right}: SO haloes with $\Delta=200\mathrm{c}$.
    The light and dark shaded areas mark the $\pm5\%$ and $\pm10\%$ levels, respectively.
	}
	\label{fig:emu_vs_fits}
\end{figure*}

\subsubsection{$\Lambda$CDM}
\label{subsubsec:other_preds_lcdm}

In Fig.~\ref{fig:emu_vs_fits}, we plot the relative difference between several popular predictions \citep[namely][]{Tinker_2008,Crocce_2010,Bhattacharya_2011,Courtin_2011,Angulo_2012,Watson_2013,Despali_2016,Bocquet_2016}, as implemented in \textsc{colossus} \citep{Diemer_2018}, and the \textsc{e-mantis} HMF at $z=0$, $z=0.5$, $z=1$, and $z=1.5$ for the P18 reference cosmology.
We consider two common choices of halo definition: FoF haloes with a linking-length $b=0.2$ widely used in the $N$-body simulation community, and SO haloes with an overdensity threshold $\Delta=200\mathrm{c}$ widely used in the galaxy cluster community.
We also compare our prediction to that of \textsc{mira-titan} emulator \citep{mira_titan_hmf_emu}.
While we only consider one external emulator in our comparison, we note that other HMF emulators are available in the literature, such as \textsc{dark emulator} \citep{dark_quest_hmf_emu} and \textsc{aemulus} \citep{aemulus_hmf_emu}.
However, to the best of our knowledge, these are restricted to other mass definitions, such as $200$ times the mean background density.
Overall, it is important to stress that all the models agree with the \textsc{e-mantis} predictions, with relative deviations of less than $10\%$ over a wide range of redshifts and masses.
The deviations only exceed the $10\%$ level for some predictions at the high-mass end, where the statistical noise related to the limited simulation volume is expected to be large due to the steepness of the mass function.
In this regime, the estimated HMF is also very sensitive to initial condition truncation and discreteness errors, finite mass resolution effects, and other systematic effects.

In the following, we discuss the comparison with the different predictions case by case.
Notice that as our emulator provides predictions with an accuracy better than $\sim5\%$ when confronted to external $N$-body simulations, then differences of the various HMF calibrated parametrizations will be indicative of their accuracy in reproducing the simulation results:
\begin{itemize}
  \item \emph{\citet{Crocce_2010} FoF mass function}:
        the agreement is at the $5\%$ level for $M_{h} \lesssim 10^{15}\,h^{-1}M_{\odot}$ at $z=0$, while it degrades to the $7\%$ for $M_{h} \lesssim3\cdot10^{14}\,h^{-1}M_{\odot}$ at $z=1$.
        In contrast, the deviation with respect to the prediction from \textsc{e-mantis} increases up to the $20\%$ level at the very high mass-end.
        We do not show the comparison at redshifts $z>1$, since these are outside the limit of validity of their fitting formula.

  \item \emph{\citet{Bhattacharya_2011} FoF mass function}:
        the agreement is at the $3\%$ level from $z=0$ to $z=1$ over a wide range of masses (up to $10^{15}\,h^{-1}M_{\odot}$ at $z=0$ and $3\cdot10^{14}\,h^{-1}M_{\odot}$ at $z=1$).
        At $z=1.5$ the fit is $\sim 5\%$ below our prediction.

  \item \emph{\citet{Courtin_2011} FoF mass function}:
        the agreement is at the $5\%$ level up to $3\cdot10^{14}\,h^{-1}M_{\odot}$ at $z=0$.
        The difference increases up $10-20\%$ between $9\cdot10^{14}\,h^{-1}M_{\odot}$ and $1.3\cdot10^{15}\,h^{-1}M_{\odot}$, while the fit is not valid beyond this mass scale.
        We only show the comparison at $z=0$, as for higher redshifts it would require a conversion of the FoF halo masses that is beyond the scope of this analysis.
  \item \emph{\citet{Angulo_2012} FoF mass function}:
        the agreement is between $2$ and $3\%$ level up to $1.5\cdot10^{15}\,h^{-1}M_{\odot}$ at $z=0$, while no fitting formula is provided at higher redshift.

  \item \emph{\citet{Watson_2013} FoF mass function}:
        at $z=0$ the HMF prediction is below that of our emulator at $\sim10\%$ level in the low-mass range and improves to $\sim5\%$ up to $10^{15}\,h^{-1}M_{\odot}$.
        At $z=0.5$ the agreement is at $3\%$ level up to $9\cdot10^{14}\,h^{-1}M_{\odot}$ and similarly for $z=1$ up $2\cdot10^{14}\,h^{-1}M_{\odot}$.
        At $z=1.5$ the agreement is $\sim3\%$ at the low-mass end and degrades beyond $10\%$ level at the high-mass end with steep slope probably related to the redshift-independence of the fitting parameters.

  \item \emph{\citet{Tinker_2008} SO mass function}:
        the agreement with the \textsc{e-mantis} prediction at $z=0$ is on average at $1\%$ level across the full mass range.
        At higher redshifts, the fit falls progressively below the emulator.
        For instance, at $z=0.5$ it drops on average at $\sim2\%$, at $z=1$ is of the order of $\sim5\%$, while at $z=1.5$ the difference amounts on average to $\sim12\%$ (ranging from $7\%$ near $10^{13}\,h^{-1}M_{\odot}$ to $20\%$ near $1.5\cdot10^{14}\,h^{-1}M_{\odot}$).
        Recent analyses of numerical simulations as well as the testing of emulators have shown that the \citet{Tinker_2008} fit is only accurate at low redshifts.
        As an example, the \textsc{mira-titan} emulator is $10\%$ above the fit, as we will discuss below.
        Moreover, \citet{aemulus_hmf_emu} have shown that while the precision of the fit is of order $\sim5\%$ at $z=0$, errors can reach very large value at $z=2$.
        \citet{dark_quest_hmf_emu} has also shown very precisely that at $z=1.48$ their simulations predict a larger mass function by $13\%$ near $10^{13}\,h^{-1}M_{\odot}$ to $18\%$ near $10^{14}\,h^{-1}M_{\odot}$, which is compatible with our finding given that the mass definition is slightly different at this redshift.

  \item \emph{\citet{Despali_2016} SO mass function}:
        the agreement at $z=0$ is at the $2\%$ level up to $2\cdot10^{14}\,h^{-1}M_{\odot}$, while at larger masses the fit strongly deviates from the emulator prediction, reaching a $25\%$ discrepancy at $10^{15}\,h^{-1}M_{\odot}$.
        This very steep increase is likely due to the more rigid functional form of the fitting function, making it difficult to accurately capture the shape of the HMF at large masses.
        At $z=0.5$ the agreement degrades to $5\%$ level up $3.5\cdot10^{14}\,h^{-1}M_{\odot}$ and increases at larger masses up to $\approx 10\%$ level.
        We can see that the agreement further degrade at the $\sim10\%$ level at $z=1$ and the $\sim15\%$ level at $z=1.5$.
        Deviations from \citet{Despali_2016} have already been found in the literature \citep[see e.g.][]{Shirasaki_2021}.

  \item \emph{\citet{Bocquet_2016} SO mass function}:
        at $z=0$ the agreement is at the $\sim8\%$ level in the low mass end (see the next item for a more detailed discussion), and improves up to $\sim3\%$ in the high mass end.
        At higher redshifts, the \textsc{Colossus} implementation of this fit gives predictions that are not in agreement with the results presented in \citet{Bocquet_2016}, with differences of up to $20-30\%$.
        Therefore, we suspect a bug in the implementation and exclude redshifts $z>0$ from this comparison.

  \item \emph{\citet{Bocquet2019} SO mass function (\textsc{mira-titan} emulator)}:
        the agreement at $z=0$ is at the $\sim8\%$ level in the low mass range $M_h\lesssim 10^{14}\,h^{-1}M_{\odot}$, it improves up to $\sim3\%$ in the mass interval $10^{14}\lesssim M_h\,[h^{-1}M_{\odot}]\lesssim 6\cdot10^{14}$ and then degrades up to $10\%$ level at higher masses.
        We find the same level of agreement at $z=0.5,1$ and $1.5$, though in the latter case we find a deviation exceeding the $10\%$ level at the very high-mass end.
        It is important to stress that the redshift evolution is in much better agreement than that found in the case of the \citet{Tinker_2008} and \citet{Despali_2016} fits.
        The $\sim 8\%$ discrepancy at small masses is consistent with the results of the comparison performed in \citet{Bocquet2019} between the $z=0$ \citet{Tinker_2008} mass function, the re-scaled \textsc{aemulus} mass function and the \textsc{mira-titan} mass function, where the authors have found that \textsc{aemulus} and \citet{Tinker_2008} mass functions are in good agreement, while the \textsc{mira-titan} is $8\%$ below.
        This is not an error of \textsc{mira-titan}, but the consequence of having a different halo mass definition.
        In fact, while our emulator is calibrated on SO halo catalogues, the \textsc{mira-titan} uses a mixture of FoF detection with SO-like masses. This also accounts for the difference at low mass between \textsc{e-mantis} and the fit from \citet{Bocquet_2016}.
        Moreover, differently from the \textsc{e-mantis} emulator, no mass resolution correction has been accounted in their training data.
\end{itemize}

\begin{figure*}
	\centering
	\begin{subfigure}{0.49\linewidth}
		\includegraphics[width=0.95\linewidth]{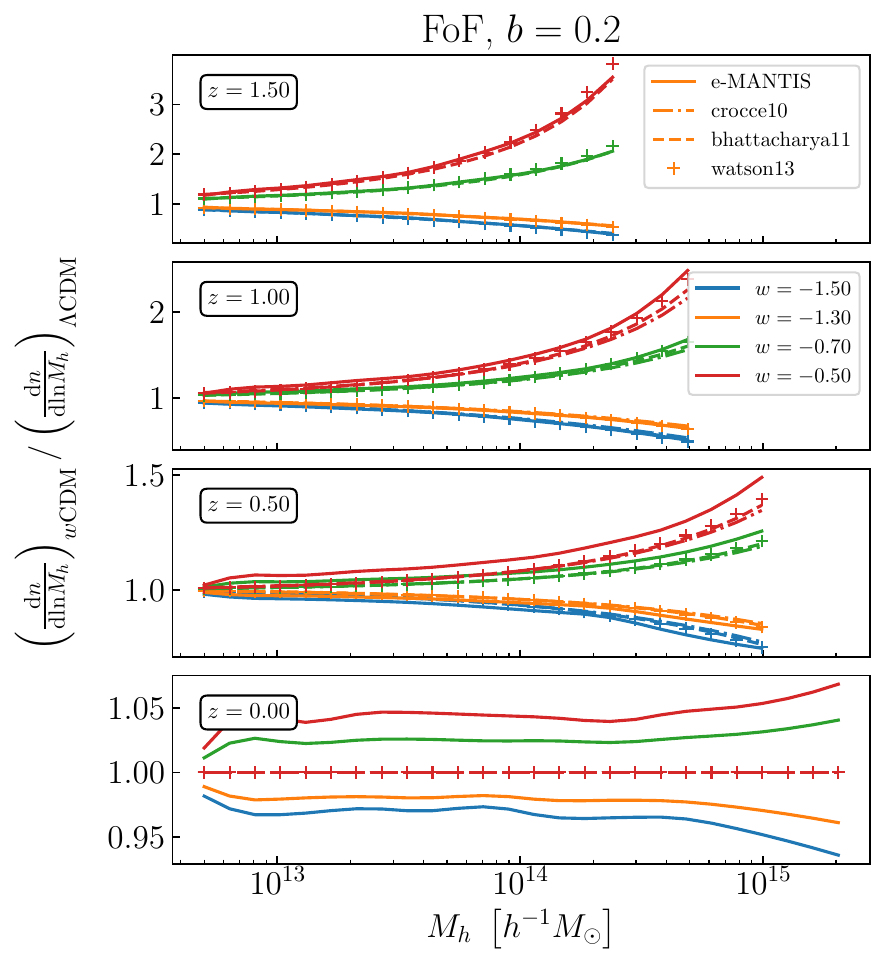}
	\end{subfigure}
	\begin{subfigure}{0.49\linewidth}
		\includegraphics[width=0.95\linewidth]{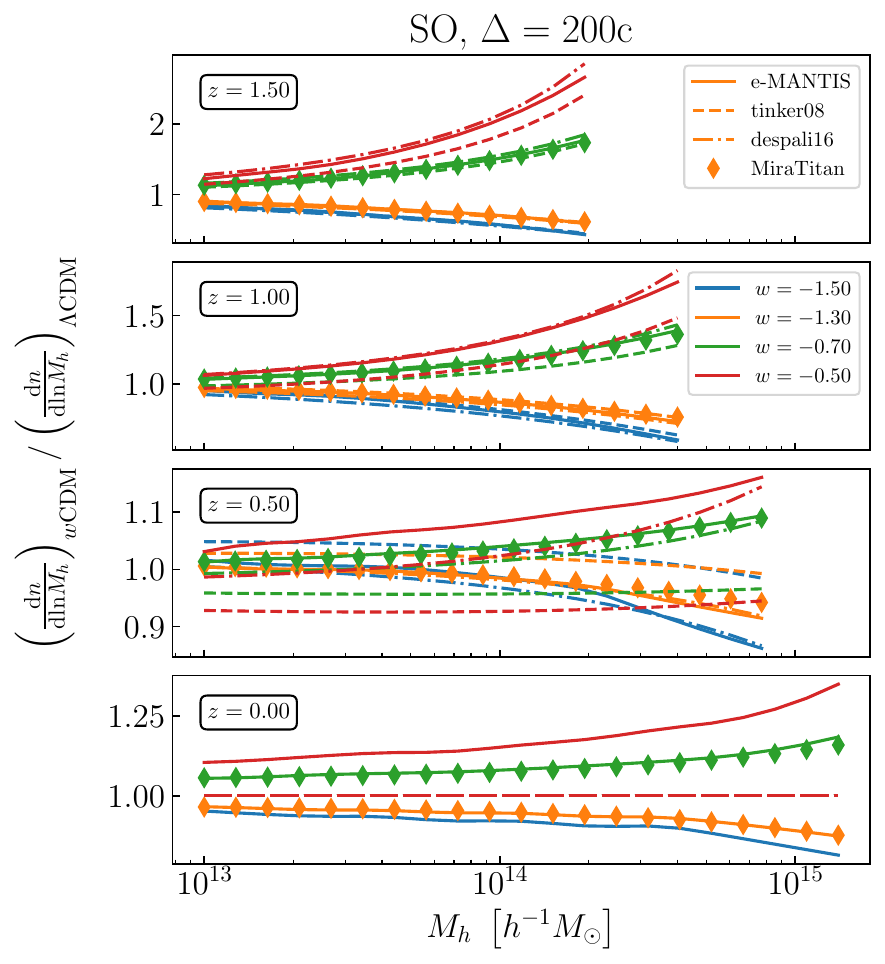}
	\end{subfigure}
	\caption{
        Ratio of the HMF in $w$CDM with respect to $\Lambda$CDM for different values of $w$ and at different redshifts, as predicted by \textsc{e-mantis} as well as multiple HMF predictions from other works.
		\textit{Left}: FoF haloes with $b=0.2$.
		\textit{Right}: SO haloes with $\Delta=200\mathrm{c}$.
        Only the dedicated emulators are able to fully capture the cosmological and redshift dependence of the HMF.
	}
	\label{fig:emu_vs_fits_w}
\end{figure*}

To conclude the comparison of the HMF predicted by parametric models and the \textsc{e-mantis} predictions in the $\Lambda$CDM case, we found an overall good agreement at $z=0$ while discrepancies become more important at higher redshifts.
While some fits are specialized to particular halo definitions \citep[e.g.][]{Bhattacharya_2011,Angulo_2012}, others are very flexible \citep[e.g.][]{Tinker_2008, Despali_2016}, which comes at the cost of precision for a given mass definition. The comparison with the predictions of the \textsc{mira-titan} emulator show differences $\sim 3-10\%$. These primarily result from the fact that \textsc{e-mantis} emulator for different overdensity thresholds has been calibrated using halo catalogues generated with a SO halo finder, while \textsc{mira-titan} relies on FoF halos with spherical masses at different overdensities. Moreover, \textsc{e-mantis} accounts for mass resolution effects on the predicted HMF.

\subsubsection{$w$CDM \& \texorpdfstring{$f(R)$}CDM}

\begin{figure*}
	\centering
	\begin{subfigure}{0.49\linewidth}
		\includegraphics[width=0.95\linewidth]{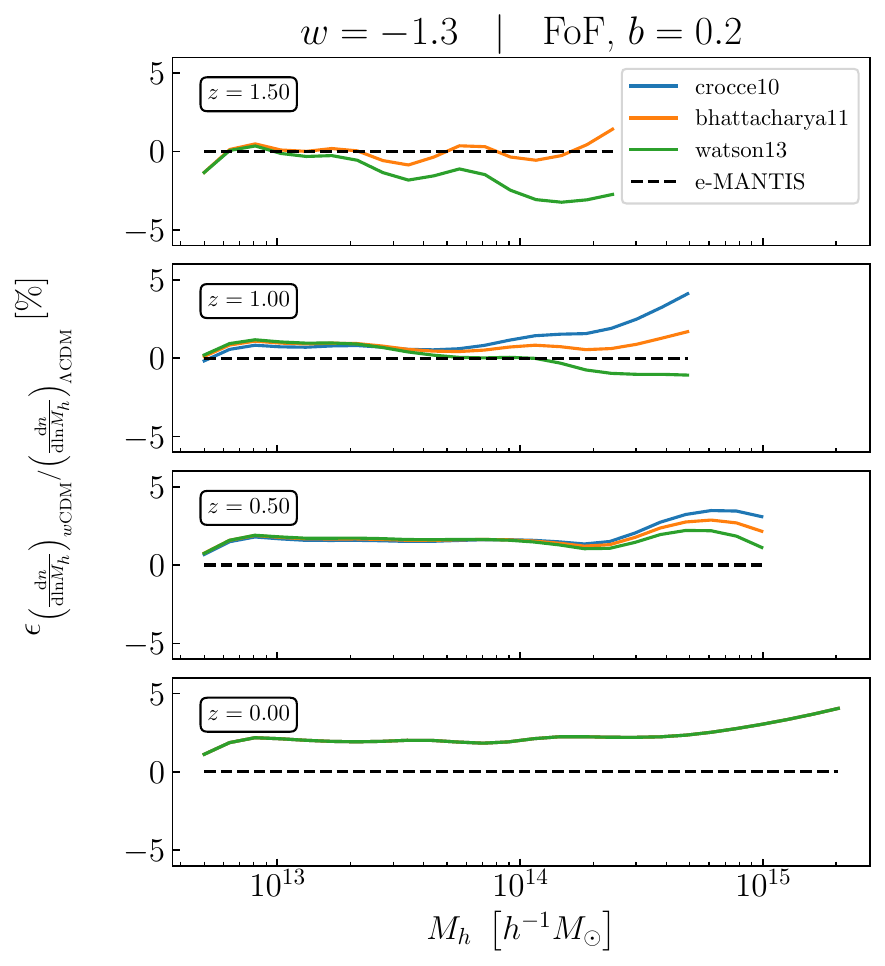}
	\end{subfigure}
	\begin{subfigure}{0.49\linewidth}
		\includegraphics[width=0.95\linewidth]{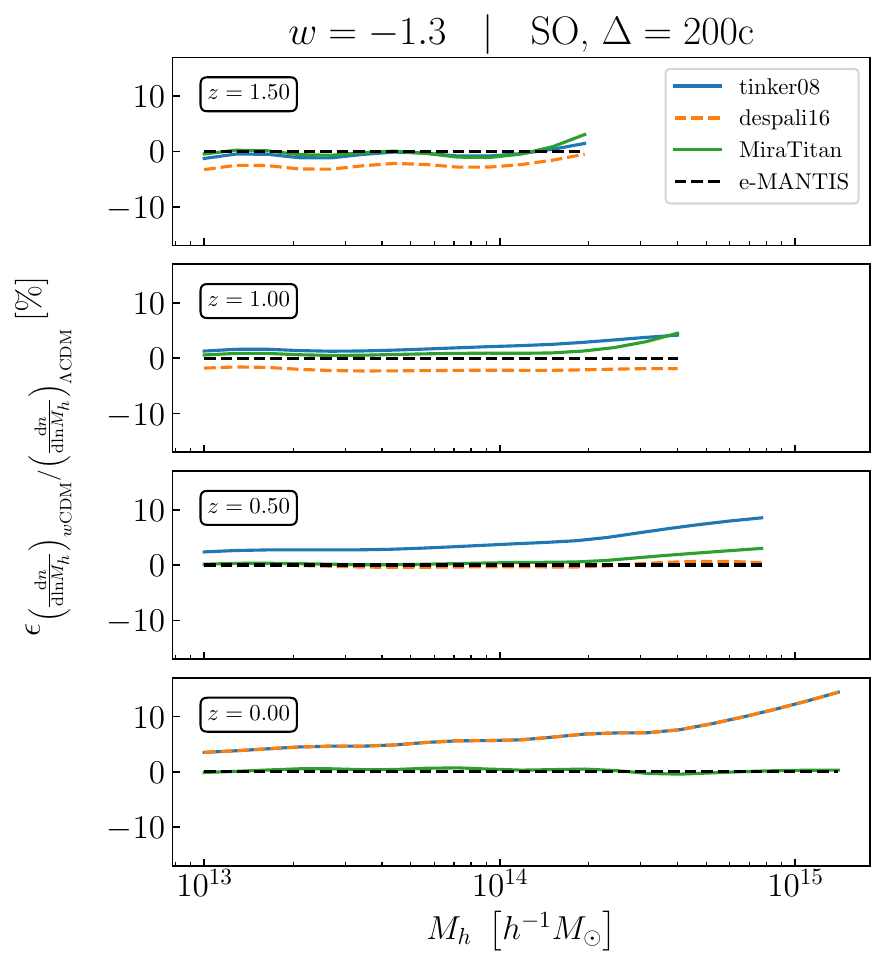}
	\end{subfigure}
	\caption{
		Relative difference between different existing predictions and \textsc{e-mantis}, for the ratio of the HMF in $w$CDM with respect to $\Lambda$CDM.
        We consider a $w$CDM model with $w=-1.3$.
		\textit{Left}: FoF haloes with $b=0.2$.
		\textit{Right}: SO haloes with $\Delta=200\mathrm{c}$.
	}
	\label{fig:emu_vs_fits_w_rel_diff}
\end{figure*}

\begin{figure}
	\centering
	\includegraphics[width=0.95\linewidth]{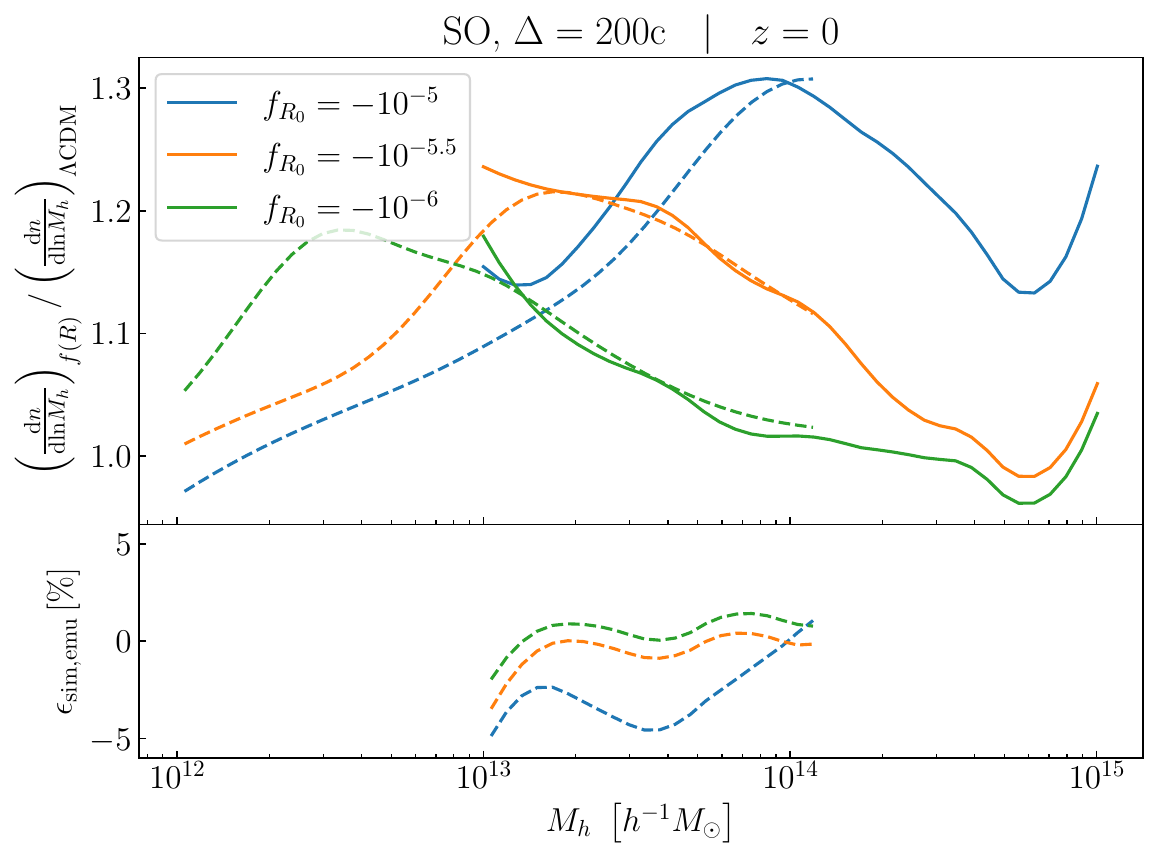}
	\caption{
    Top panel: ratio of the HMF in $f(R)$ gravity with respect to $\Lambda$CDM for different values of $f_{R_{0}}$, as predicted by \textsc{e-mantis} (solid lines) and the emulator of~\citet{forge_hmf_emu} (dashed lines), at $z=0$ and for SO haloes with $\Delta = 200\mathrm{c}$.
    Bottom panel: relative difference, in the common mass range, between the predictions from \citet{forge_hmf_emu} and \textsc{e-mantis}.
    Both emulators are compatible at the few percent level, and have complementary mass ranges.
	}
	\label{fig:emu_vs_forge_fR0}
\end{figure}

We now compare the predictions of the calibrated parametrizations of the HMF as well as those from other emulators to those of the \textsc{e-mantis} emulator in the case of $w$CDM and $f(R)$CDM cosmologies respectively.
These are shown in Fig.~\ref{fig:emu_vs_fits_w}, Fig~\ref{fig:emu_vs_fits_w_rel_diff} and Fig.~\ref{fig:emu_vs_forge_fR0}, where we plot the boost, that is the ratio of the HMF predicted in the non-standard cosmological scenario with respect to the $\Lambda$CDM case, as well as the corresponding relative difference with respect to \textsc{e-mantis}.

In Fig.~\ref{fig:emu_vs_fits_w}, we show the HMF boost for the $w$CDM scenario as predicted by \textsc{e-mantis} against other HMF predictions. In particular, we consider multiple values of the dark energy equation of state, ranging from $w=-1.5$ to $w=-0.5$, while the other cosmological parameters are fixed to the values of the reference P$18$ cosmological model. We focus on two different halo definitions: SO haloes with $\Delta=200\mathrm{c}$ and FoF haloes with $b=0.2$.
At $z=0$, the parametric fits fail to capture the dependence of the boost on $w$. This is because their predictions only rely on the dependence of the HMF on the linear matter power spectrum, which at $z=0$ is identical for all values of $w$, since we have fixed $\Omega_{\rm m}$ and $\sigma_8^{\rm GR}$. However, deviations from a universal HMF are expected, since the linear growth rate $f=\mathrm{d}\ln{D_+}/\mathrm{d}\ln{a}$ (where $D_+$ is the linear growth factor) does depend on the equation of state of dark energy $w$ \citep{Courtin_2011,Ondaro-Mallea_2022}. Hence, in the case of the parametric fits, this translates into an error on the predicted HMF boost of up to $\sim5\%$ for FoF haloes, and up to $\sim20\%$ for SO haloes. We can see that this is not the case of the \textsc{mira-titan} emulator, for which predictions are in very good agreement with \textsc{e-mantis} at all redshifts, with only deviations of a few per cent at the largest masses. Indeed, the emulators are more efficient at capturing the impact of the dark energy equation of state on the HMF. Nevertheless, given the smaller range of values of $w$ covered by \textsc{mira-titan} we were not able to compare the HMF predictions for $w=-0.5$ and $w=-1.5$. It is worth noticing that at $z>0$ the parametric fits perform better than at $z=0$, since they capture part of the cosmological dependence of the HMF caused by variations in the linear matter power spectrum.

In Fig.~\ref{fig:emu_vs_fits_w_rel_diff}, we plot the relative difference between the various predictions of the $w$CDM boost with respect to \textsc{e-mantis} at different redshifts, again for FoF haloes with $b=0.2$ and SO haloes with $\Delta=200\mathrm{c}$.
As the trends are similar for all $w$CDM models, and to improve the readability of the graph, we focus on a single dark energy model with $w=-1.3$.
At $z=0$, we find $\sim 2-4$\% errors in the case of FoF haloes, and $3-15$\% errors in the case of SO one.
At $z>0$, in both the FoF and SO cases, the resulting error is of order $\pm2-3\%$ for most of the mass range, while at high masses it can reach the $5\%$ level for FoF haloes and the $10\%$ level for SO haloes.

We now turn to the $f(R)$CDM case.
Analytical predictions of the mass function in modified gravity is a notoriously difficult problem.
Using standard parametric fits, such as those previously discussed, with modified linear power spectra and with a fixed critical density $\delta_c$ from the spherical collapse model, fails to describe the HMF from simulations of modified gravity models.
Accounting for the screening mechanism is highly non-trivial.
Many attempts exist to predict the mass function in $f(R)$ gravity using various formulation of the excursion set theory, with a critical density threshold that depends on mass and possibly environment \citep{Li_2012_EST,Lam_2012,Lombriser_2013,Kopp_2013,Difu_2015,Achitouv_2016,vBB_2018,Gupta_2022}.
The implementation of all these complex models is far beyond the scope of the work presented here.
Hence, we have opted to compare the predictions for the SO mass function boost at $z=0$ from the \textsc{e-mantis} emulator to the predictions of the $f(R)$ gravity emulator built by \citet{Ruan_2022}.
This emulator is based on the $N$-body simulation suite from \citet{Arnold2021}, whose simulations have been run with a modified gravity version of the \textsc{arepo} code \citep{Springel_2010, Arnold_2019}. 
It is complementary to our emulator since it targets the masses range $10^{12}< M_h[h^{-1}M_{\odot}]<10^{14}$, while we focus on the mass range of groups and clusters $10^{13}<M_h[h^{-1}M_{\odot}]<10^{15}$. 

In Fig.~\ref{fig:emu_vs_forge_fR0}, we plot the boost predicted by the two emulators for $f(R)$CDM models with $|f_{R_0}|=10^{-6},10^{-5.5}$ and $10^{-5}$ at $z=0$, where we have fixed the other cosmological parameter values to those of the \textsc{elephant} simulations \citep{Cautun2018}.
We can see that the shape of the boost is similar for all $f_{R_0}$ values.
In the common range of masses, there is an excellent agreement {for the HMF boost predicted by} the two emulators at the $2\%$ level for F$6$ and F$5.5$, while deviations are smaller than $5\%$ for F$5$.
Given that the emulators have been trained on different simulation datasets, using different $N$-body and halo finder codes, the $5\%$ difference is a good achievement.
To conclude, \corr{in terms of the HMF boost}, the emulators are compatible at the few percent level and they span a complementary mass range.

\section{Conclusions}
\label{sec:conclusion}

We have presented a novel emulator of the halo mass function implemented in the framework of the \textsc{e-mantis} emulator, originally developed to provide accurate predictions of the non-linear matter power spectrum of $f(R)$CDM cosmologies \citep{emantis_ps_boost}.
Not only we have implemented a new functionality, we have also extended the cosmological parameter space for which \textsc{e-mantis} can provide accurate predictions, and which now include the case of $w$CDM models.
For this purpose, we have realized a large suite of $N$-body simulations using a Latin hyper cube sampling of the six dimensional parameter space consisting of the standard cosmological parameters $\Omega_m$, $\sigma_8$, $h$, $n_s$, and $\Omega_b$ and $f_{R_0}$ for $f(R)$CDM, and $w$ for $w$CDM models.
We consider a large range for the non-standard parameters: $f_{R_0}\in[-7, -4]$ and $w\in[-1.5, -0.5]$.
In particular, for each model the simulations spans $80$ different combination of parameters for different simulation volume, mass resolution and random phase of the initial conditions.
The total number of simulations in this extended \textsc{e-mantis} simulation suite exceeds 10000.

We have generated halo catalogues for each of the simulations in the suite using two different halo finders.
Hence, contrary to other emulators presented in the literature \citep[see eg.][]{aemulus_hmf_emu, dark_quest_hmf_emu, mira_titan_hmf_emu}, the \textsc{e-mantis} emulator of the HMF is capable of providing accurate and fast predictions in the case of FoF halos \corr{with linking lenght} $b=0.2$ as well as SO halos with overdensities values compatible with those of observable mass definitions of galaxy clusters $\Delta=200\mathrm{c},500\mathrm{c},1000\mathrm{c}$.
In the latter case, we have detected SO halos with a detection threshold of $\Delta=200\mathrm{c}$ and then computed the halo masses at $\Delta=500\mathrm{c}$ and $1000\mathrm{c}$ from the mass profile. 

We have built HMF training datasets using a B-spline basis decomposition of the HMFs estimated from the simulation suite.
This is particularly important if one aims to obtain accurate predictions of the HMF for modified gravity models, since traditional fitting formulae do not perform as well as in standard GR base scenarios \citep[see e.g.][]{Gupta_2022}.
Indeed, we find that the B-spline decomposition is able to capture the shape of the HMFs with per cent level accuracy at low masses, while at large masses the performance of the fit is only limited by the statistical noise of the simulation data.
We use a Gaussian process regression to emulate the cosmological dependence of the B-spline coefficients, while we use a linear interpolation of the logarithm of the HMF as a function of scale factor to capture the redshift dependence.
The final emulator is able to predict the HMF for the different halo definitions described above in the redshift range $0 \leq z \leq 1.5$ and over the range of masses $10^{13} \lesssim M_{h}[h^{-1}M_{\odot}]\lesssim 10^{15}$ for which the emulator has been designed, both for $w$CDM and $f(R)$CDM models. 

We have performed a thorough analysis to assess the accuracy of the emulator predictions.
In particular, the results of the leave-one-out tests show that the emulator achieves a few per cent level accuracy in the low mass end of the HMF, where the error budget is dominated by the emulation itself.
In the higher mass end, the emulation error is dominated by the statistical noise in the training HMF simulation data.
Overall, the results indicate better than a few percent accuracy over a wide range of halo masses.
As an example, the errors on the HMF prediction for SO halos at $z=0$ in the case of $w$CDM models are at $\sim 1.5\%$ level up to $M_{200c}=2\cdot10^{14}\,h^{-1}M_\odot$.
\corr{
  The comparison with the HMF estimated from the large volume high-resolution Uchuu simulation shows an agreement better than $5\%$ for all masses and redshifts considered, and further validates the mass resolution correction that we apply to our training data.
  In terms of the cosmological dependence of the HMF boost, i.e. the ratio between the HMF in an alternative cosmology with respect to $\Lambda$CDM, we find an agreement better than $5\%$ over a wide range of masses with the measurements from the RayGal, i.e. in $w$CDM, and Elephant, i.e. in $f(R)$CDM, simulations.
}

This is a noticeable result, especially in the case of the Uchuu and Elephant simulations, which were realized using different halo finders than those we have used for the \textsc{e-mantis} simulation suite.
Moreover, the \textsc{e-mantis} emulator provides an estimate of its own errors, thus enabling the possibility to propagate such model prediction uncertainties in the cosmological parameter inference analysis.
We have tested the predictions against those from existing fitting formula of the HMFs as well as publicly available emulators.
We show that only emulators can account for departure from the universality of the HMF.
Ignoring such deviations can lead up to $25\%$ errors on the SO mass function at the high-mass end.
We find a good agreement with the predictions for the \textsc{mira-titan} emulator in the $w$CDM case.
The comparison with the HMF emulator for $f(R)$CDM models from \citet{forge_hmf_emu}, which targets the mass interval $10^{12}\lesssim M_{h}[h^{-1}M_{\odot}] \lesssim 10^{14}$, shows an agreement better than $5\%$, \corr{in terms of the HMF boost}, in the common mass range for different values of $f_{R_0}$. 

It is worth remarking that the \textsc{e-mantis} emulator is calibrated on dark matter only simulations and consequently does not include the effect of baryonic physics. 
The feedback caused by baryons on the matter distribution, can have an impact on the HMF at the $10-20\%$ level \citep{2012MNRAS.423.2279C,Bocquet_2016, Castro2021}.
Such effects need to be included in theoretical predictions of the HMF.
However, hydrodynamical simulations including baryonic physics are computationally orders of magnitudes more expensive than dark matter only simulations. Consequently, building an emulator based on such simulations remains very challenging. Nevertheless, recent studies have shown that in the case of dynamical dark energy scenarios (which include the case of $w$CDM models), it is possible to disentangle the impact of baryons on the HMF from that of the cosmological model with per cent level accuracy \citep{Pfeifer_2020}.
Therefore, it should be possible to include the effect of baryons on the HMF as a post-processing step.
Similarly, \citet{Arnold_2019} have shown that the impact of baryons on the HMF in $f(R)$ gravity can be estimated from hydrodynamical simulations of $\Lambda$CDM.
Hence, as a first approximation, a $\Lambda$CDM baryonic correction to the HMF can also be applied to a dark matter only predictions for $f(R)$CDM models.

The calculation time of the HMF with \textsc{e-mantis} emulator is of order $\sim 10$ ms, while it combines accuracy and the possibility of exploring a wide range of cosmologies.
Moreover, the emulator can be easily implemented in the standard Markov Chain Monte Carlo likelihood analysis pipelines.
As such, it provides a useful tool to constrain alternative dark energy models using galaxy cluster abundance measurements from the new generation of galaxy cluster surveys.

\begin{acknowledgements}

This project has benefited from computational time and data storage resources allocated by GENCI (Grand Équipement National de Calcul Intensif) at TGCC (Très Grand Centre de Calcul) for the projects 2021-A0110402287, 2022-A0130402287 and 2023-A0150402287 on the Joliot Curie supercomputer's ROME and SKYLAKE partitions.
We thank the Euclid Consortium for the “Sponsor PhD Grant” of ISC.
\corrtwo{YR acknowledges support from ANR ProGraceRay (ANR-23-CE31-0010).}
We thank Baojiu Li for sharing the \textsc{ecosmog} simulation code.
We thank Razi Sheikholeslami for sharing the PLHS \textsc{matlab} code, which served as validation for our own \textsc{python} implementation.
We thank Suhani Gupta for sharing the Elephant HMF data used in Figure~\ref{fig:emu_vs_sim_elephant}.
We are also grateful to Cheng-Zong Ruan et Baojiu Li for providing the predictions of the boost of $f(R)$CDM models with the emulator presented in \citet{Ruan_2022}.
The plots presented in this paper were produced with the \textsc{matplotlib}~\citep{matplotlib} package.
A significant amount of the numerical computations were carried out using the \textsc{scipy}~\citep{scipy}, \textsc{numpy}~\citep{numpy} and \textsc{scikit-learn}~\citep{scikit-learn} packages.

\end{acknowledgements}

\bibliographystyle{aa} 
\bibliography{main}

\onecolumn

\begin{appendix}

\section{Sampling of the cosmological parameter space}
\label{app:LHS}

\begin{figure}[!h]
  \centering
  \includegraphics[width=0.95\linewidth]{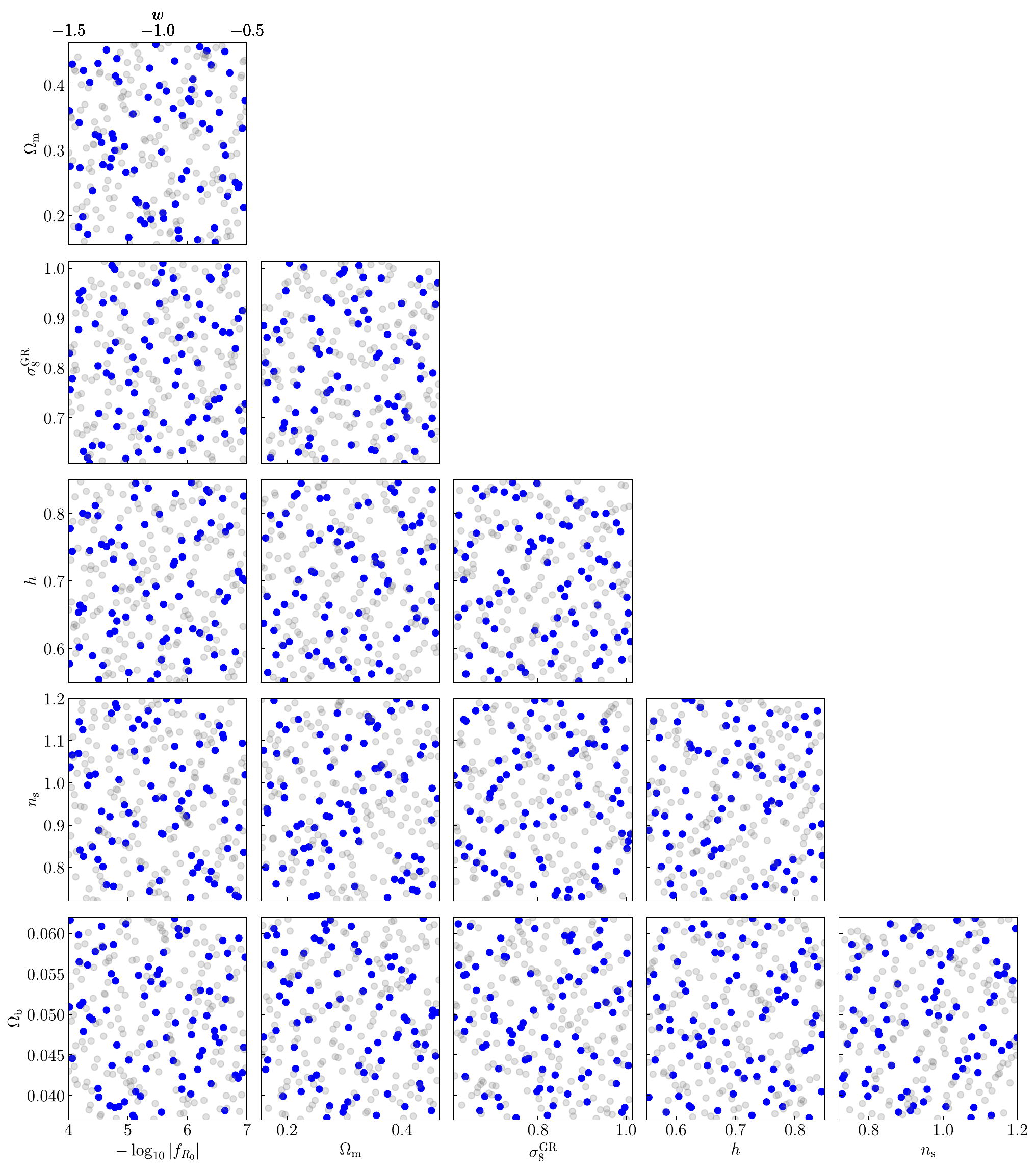}
  \caption{
    Distribution of cosmological model parameter values of the $w$CDM and $f(R)$CDM emulators.
    The blue points correspond to the $80$ training models of the emulators presented in this paper.
    The light grey points correspond to the remaining $176$ models of the quasi-PLHS that may be used for a future extension of the simulation suite.
    Notice that the $w$ and $\log_{10}\left|f_{R_0}\right|$ parameters share the same distribution of points, but rescaled to their respective range of values.
  }
  \label{fig:LHS}
\end{figure}

\FloatBarrier

\section{Numerical convergence}
\label{app:numerical_convergence}

\begin{figure}
  \centering
  \includegraphics[width=0.95\linewidth]{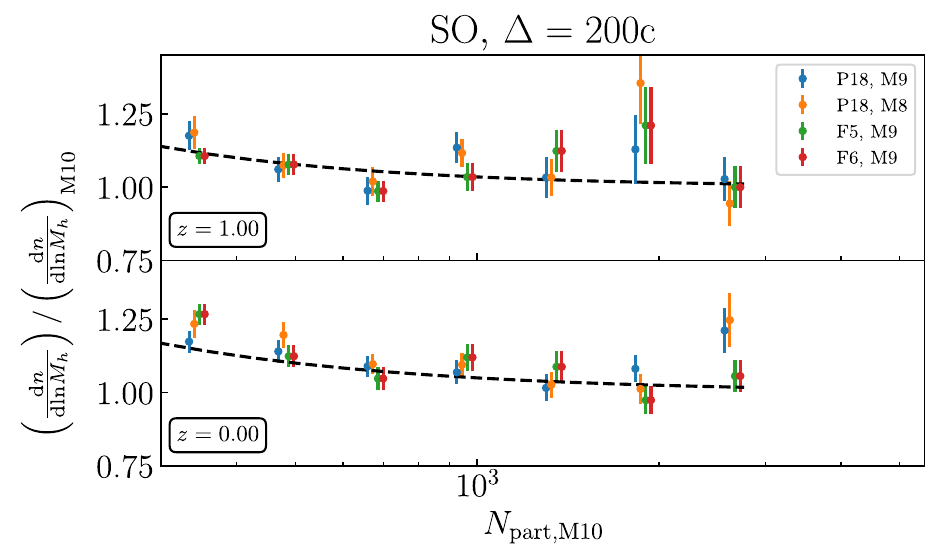}
  \caption{
    Ratio of the $\mathrm{L}164\_\mathrm{M}9$ HMF, for the cosmological models P$18$, F$5$ and F$6$, and the $\mathrm{L}164\_\mathrm{M}8$ HMF, for the model P$18$, with respect to the $\mathrm{L}164\_\mathrm{M}10$ HMF, in the case of SO haloes with $\Delta=200\mathrm{c}$ at $z=1$ and $z=0$.
    For visual clarity, the $x$ coordinates have been slightly displaced.
    The dashed lines correspond to the fit performed in Sect.~\ref{subsec:mass_res_corr}, which is used to correct the training data of the emulator from the $\mathrm{L}328\_\mathrm{M}10$ simulations.
    We do not detect any statistically significant deviations from this fit, due to the effect of $f(R)$ gravity or the higher M$8$ resolution.
  }
  \label{fig:mass_res_corr_conv_and_fR_test}
\end{figure}

In Sect.~\ref{subsec:mass_res_corr}, we have computed a mass resolution correction for the emulator $\mathrm{L}328\_\mathrm{M}10$ simulations using the higher resolution $\mathrm{L}328\_\mathrm{M}9\_\mathrm{wcdm}$ simulation set.
This correction is used to upscale the low mass end of the emulator HMF training data from the M$10$ resolution to the M$9$  resolution.
However, it is not guaranteed that the M$9$ simulations have fully converged for the mass range of interest.
In order to test this, we use a set of simulations using a box-size of $L_\mathrm{box}=164.0625\,h^{-1}\mathrm{Mpc}$ and with different number of $N$-body particles: $256^{3}$, $512^{3}$ and $1024^{3}$ corresponding to the resolutions M$10$, M$9$ and M$8$ respectively.
The initial conditions from the M$10$ and M$9$ simulations are degraded versions of the M$8$ initial conditions.
These simulations use the P$18$ cosmological model.
We compute the ratio of the M$9$ and M$8$ HMF to the M$10$ HMF.
We estimate the corresponding errors using $4^{3}$ spatial jackknife sub-volumes, in order to account for the correlation between the different simulations.
These ratios are shown in Fig.~\ref{fig:mass_res_corr_conv_and_fR_test}, along with the fitted correction used for the emulator training data, for SO haloes with $\Delta=200\mathrm{c}$ at $z=1$ and $z=0$.
We remind that the emulator only uses data for SO haloes with a number of particles larger than $N_{\mathrm{part}, M10}=300$.
We find that the in this mass range, the M$9$ HMF is in good agreement with the measurements from M$8$.
We cannot see any statistically significant deviations from our correction at higher resolution.
Therefore, the HMF from the M$9$ simulations has properly converged in the mass range of interest and no further resolution correction is required.
We obtain the same conclusion for all the halo definitions and redshifts relevant for our emulator.

The correction computed in Sect.~\ref{subsec:mass_res_corr} has been obtained from $w$CDM simulations.
We want to validate that it is still valid in the case of $f(R)$ gravity.
We use HMF measurements from simulations using the $\mathrm{L}164\_\mathrm{M}9$ simulations box, for two different values of the $f_{R_{0}}$ parameter: $f_{R_{0}}=-10^{-5}$ (F$5$) and $f_{R_{0}}=-10^{-6}$ (F$6$).
Figure~\ref{fig:mass_res_corr_conv_and_fR_test} shows that, in the case of SO haloes with $\Delta=200{\rm c}$, the HMF ratio measured from the F$5$ and F$6$ simulations is compatible the with correction fitted from the $w$CDM simulations.
We find that this result holds for the other halo definitions and redshifts considered in this work.
We can therefore safely apply such correction to the $f(R)$CDM emulator training data.

\FloatBarrier

\end{appendix}

\end{document}